\documentclass[12pt]{article} 

\usepackage{epsfig} 
\usepackage{amsbsy}  
 
\newlength{\largfig}
\def\m{m_t} 
\def\Tr#1{{\rm Tr}\left(#1\right)}

\def\ds#1{#1\kern-1ex\hbox{/}} 
\def\sl#1{#1\kern-1ex\hbox{/}} 
\def\dsh{h\kern-1.2ex /}

\def\beq{\begin{equation}} 
\def\eeq{\end{equation}} 
\def\eq{\beq\eeq} 
\def\beqn{\begin{eqnarray}} 
\def\eeqn{\end{eqnarray}} 
 
\def\lq{\left[} 
\def\rq{\right]} 
\def\rg{\right\}} 
\def\lg{\left\{} 
\def\({\left(} 
\def\){\right)} 
 
\def\ba{\begin{eqnarray}} 
\def\ea{\end{eqnarray}} 
\def\bq{\begin{equation}} 
\def\eq{\end{equation}} 
\def\lsim{\mathrel{\raisebox{-.6ex}{$\stackrel{\textstyle<}{\sim}$}}} 
\def\gsim{\mathrel{\raisebox{-.6ex}{$\stackrel{\textstyle>}{\sim}$}}} 
\def\sla#1{\ifmmode%
\setbox0=\hbox{$#1$}%
\setbox1=\hbox to\wd0{\hss$/$\hss}\else%
\setbox0=\hbox{#1}%
\setbox1=\hbox to\wd0{\hss/\hss}\fi%
#1\hskip-\wd0\box1 } 
 
\def\detw{{\rm det}{\cal Q}_2} 
\def\detx{{\rm det}{\cal Q}_3} 
\def\detxt{ {\detx\over q_1\cdot q_2} }

\def\asb{{}\ifmmode \bar{\alpha}_s \else $\bar{\alpha}_s$\fi} 
\def \as   {\ifmmode \alpha_s \else $\alpha_s$ \fi}

\def\al{\alpha}

\def\bt{\beta} 
\def\ga{\gamma} 
 
\def\de{\delta} 
 
\def\th{\theta}

\def\so3#1{\,{\rm S}_{1,\,3}\left(#1 \right)} 
\def\st2#1{\,{\rm S}_{2,\,2}\left(#1 \right)} 

\def\Re{\mathop{\rm Re}}

\newskip\humongous \humongous=0pt plus 1000pt minus 1000pt

\newif\ifdtup

\catcode`@=12 
 
\catcode`\@=11 
 
\@addtoreset{equation}{section} 

\def\theequation{\thesection.\arabic{equation}} 
 
\def\@normalsize{\@setsize\normalsize{15pt}\xiipt\@xiipt 
\abovedisplayskip 14pt plus3pt minus3pt%
\belowdisplayskip \abovedisplayskip 
\abovedisplayshortskip \z@ plus3pt%
\belowdisplayshortskip 7pt plus3.5pt minus0pt} 
 
\def\small{\@setsize\small{13.6pt}\xipt\@xipt 
\abovedisplayskip 13pt plus3pt minus3pt%
\belowdisplayskip \abovedisplayskip 
\abovedisplayshortskip \z@ plus3pt%
\belowdisplayshortskip 7pt plus3.5pt minus0pt 
\def\@listi{\parsep 4.5pt plus 2pt minus 1pt 
     \itemsep \parsep 
     \topsep 9pt plus 3pt minus 3pt}} 
 
\@twosidetrue

\catcode`\@=11  
\def\section{\@startsection{section}{1}{\z@}{3.5ex plus 1ex minus 
   .2ex}{2.3ex plus .2ex}{\large\bf}}

\def\thesection{\arabic{section}} 
\def\thesubsection{\arabic{section}.\arabic{subsection}} 
\def\thesubsubsection{\arabic{section}.\arabic{subsection}.\arabic{subsubsection}} 
 
\def\appendix{\setcounter{section}{0} 
 \def\thesection{\Alph{section}} 
 \def\theequation{\Alph{section}.\arabic{equation}} 
 \def\thesubsection{\Alph{section}.\arabic{subsection}} 
\def\thesubsubsection{\Alph{section}.\arabic{subsection}.\arabic{subsubsection}} 
 
\def\section{\@startsection{section}{1}{\z@}{3.5ex plus 1ex minus 
   .2ex}{2.3ex plus .2ex}{\large\bf}} 
}

\newcommand{\ccaption}[2]{ 
  \begin{center} 
    \parbox{0.85\textwidth}{ 
      \caption[#1]{\small\it {#2}}} 
  \end{center}    } 
 
\def \ep{\epsilon} 
\def \eps{\epsilon} 
\def \to   {\mbox{$\rightarrow$}}

\newcount\minutes 
\newcount\scratch 
 
\def\timestamp{%
\scratch=\time 
\divide\scratch by 60 
\edef\hours{\the\scratch} 
\multiply\scratch by 60 
\minutes=\time 
\advance\minutes by -\scratch 
---$\,$\hours:\null 
\ifnum\minutes< 10 0\fi 
\the\minutes}

\begin{document} 
\begin{titlepage} 
\nopagebreak 
{\flushright{ 
        \begin{minipage}{5cm} 
         MADPH 01-1235 \\ 
	 BNL-HET-01/28\\ 
         MSUHEP-10709\\	 
         DFTT 19/2001\\ 
        {\tt hep-ph/0108030}\hfill \\ 
        \end{minipage}        } 
 
} 
\vfill 
\begin{center} 
{\LARGE \bf \sc 
 \baselineskip 0.9cm 
Gluon-fusion contributions to $H+2$~jet production 
           
} 
\vskip 0.5cm  
{\large   
V.~Del Duca$^a$, W.~Kilgore$^b$, C.~Oleari$^c$, C.~Schmidt$^d$ and  
D.~Zeppenfeld$^c$ 
}   
\vskip .2cm  
{$^{(a)}$ {\it I.N.F.N., Sezione di Torino 
via P.~Giuria, 1 - 10125 Torino, Italy}}\\  
{$^{(b)}$ {\it Physics Department, 
  Brookhaven National Laboratory,  
  Upton, New York 11973, U.S.A.}}\\ 
{$^{(c)}$ {\it Department of Physics, University of Wisconsin, Madison, WI 
53706, U.S.A. }}\\   
{$^{(d)}$ {\it Department of Physics and Astronomy, 
Michigan State University, 
East Lansing, MI 48824, U.S.A.}}\\

\vskip 
1.3cm     
\end{center} 
 
\nopagebreak 
\begin{abstract}
Real emission corrections to Higgs production via gluon fusion, at order 
$\alpha_s^4$, lead to a  Higgs plus two-jet final state.
We present the calculation of these
scattering amplitudes, as induced by top-quark triangle-, box- and
pentagon-loop diagrams. These diagrams are evaluated analytically for 
arbitrary top mass $m_t$.
We study the renormalization and factorization scale-dependence of the 
resulting $H+2$~jet cross section, and discuss phenomenologically 
important distributions at the LHC. The gluon fusion results
are compared to expectations for weak-boson fusion cross sections.
\end{abstract} 
\vfill 
\vfill 
\end{titlepage} 
\newpage

\section{Introduction} 

Gluon fusion and weak-boson fusion are expected to be the most copious
sources of Higgs bosons in $pp$-collisions at the Large Hadron 
Collider (LHC) at  CERN. Beyond representing the most promising discovery 
processes~\cite{CMS,ATLAS}, these two production modes are also expected
to provide a wealth of information on Higgs couplings to
gauge bosons and fermions~\cite{Zeppenfeld:2000td}. The extraction of Higgs
boson couplings, in particular, requires precise predictions of production
cross sections.
 
Next-to-leading order (NLO) QCD corrections to the inclusive gluon-fusion  
cross section are known to be large, leading to a $K$-factor close to 
two~\cite{HggNLO}. Because the 
lowest order process is loop induced, a full NNLO calculation would entail 
a three-loop evaluation, which presently is not feasible. In the 
intermediate Higgs mass range, which is favored by electroweak precision
data~\cite{LEPEWWG}, the Higgs boson mass $m_H$ is small compared to the 
top-quark pair threshold and the large $\m$ limit promises to be an 
adequate approximation.  
Consequently, present efforts on a NNLO calculation of the inclusive  
gluon-fusion cross section 
concentrate on the $\m\,\to\,\infty$ limit, in which the task reduces to an  
effective two-loop calculation~\cite{H2loop}.  
In order to assess the validity 
of this approximation, gluon-fusion cross-section calculations, which include 
all finite $\m$ corrections, are needed.  
Of particular interest are  
phase space regions where one or several of the kinematical invariants  
are of the order of, or exceed, the top-quark mass, i.e.\ regions of 
large Higgs boson or jet transverse momenta, or regions where dijet invariant  
masses become large. For larger Higgs boson masses, top-mass corrections 
become important and a full calculation of $H+2$~jet production is needed.
 
A key component of the program to measure Higgs boson couplings at the LHC is
the weak-boson fusion (WBF) process, $qq\to qqH$ via $t$-channel $W$ or $Z$
exchange, characterized by two forward quark jets~\cite{Zeppenfeld:2000td}.
QCD radiative corrections to WBF are known to be small~\cite{WBF_NLO} and,
hence, this process promises small systematic errors. $H+2$~jet production
via gluon fusion, while part of the inclusive Higgs signal, constitutes a
background when trying to isolate the $HWW$ and $HZZ$ couplings responsible
for the WBF process. A precise description of this background is needed in
order to separate the two major sources of $H+2$~jet events: one needs to
find characteristic distributions which distinguish the weak boson fusion
process from gluon fusion. One such feature is the typical large invariant
mass of the two quark jets in WBF.  A priori, this large kinematic invariant,
$m_{jj}^2\gg 4\m^2$, invalidates the heavy top approximation and requires a
full evaluation of all top-mass effects. We will find, however, that even in
this phase-space region the large $\m$ limit works extremely well, provided
that jet transverse momenta remain small compared to $\m$.

In a previous letter~\cite{DKOSZ} we presented first results of our 
evaluation of the real-emission corrections to gluon fusion which lead 
to $H+2$~parton final states, at order $\alpha_s^4$. The contributing 
subprocesses include quark-quark scattering which involves top-quark 
triangles, quark-gluon scattering processes which are mediated by 
top-quark triangles and boxes, and gluon scattering which requires
pentagon diagrams in addition. The purpose of this 
paper is to provide details of our calculation and to give a more complete
discussion of its phenomenological implications. 
In Section~\ref{sec:calculation}, we start with a brief overview of the 
calculation. Full expressions for the quark-quark and the 
quark-gluon scattering amplitudes are given in Section~\ref{sec:amp}. 
Expressions for the $gg\to ggH$ amplitudes, which were obtained by
symbolic manipulation, are too long to be given explicitly. Instead we 
describe the details of the calculational procedure in 
Section~\ref{sec:gg_ggH}. The matrix elements for all subprocesses have 
been checked both analytically and numerically. The most important of these 
tests are described in Section~\ref{sec:checks}. 
We then turn to numerical results, in particular
to implications for LHC phenomenology. In Section~\ref{sec:pheno}, we
first compare overall $H+2$~jet cross sections from weak-boson fusion and
from gluon fusion and determine the subprocess decomposition of the latter.
QCD uncertainties are assessed via a discussion of the scale dependence
(renormalization and factorization) of
our results. We investigate various distributions, searching for
characteristic
differences between gluon fusion and WBF. Our final conclusions are given 
in Section~\ref{sec:concl}. 

A number of technical details are collected in the Appendixes. 
Scalar integrals, in particular the evaluation of scalar five-point functions,
are discussed in Appendix~\ref{app:CDE}. Appendix~\ref{app:identities} gives
useful relations among Passarino-Veltman $C_{ij}$ and $D_{ij}$ functions.
Finally, in Appendixes~\ref{app:triangle}, \ref{app:boxes}, and 
\ref{app:pentagons}, we provide expressions for the color decomposition 
and the tensor integrals encountered in triangle, box and pentagon
graphs.

\section{Outline of the calculation} 
\label{sec:calculation} 
 
\begin{figure}[thb] 
\centerline{ 
\epsfig{figure=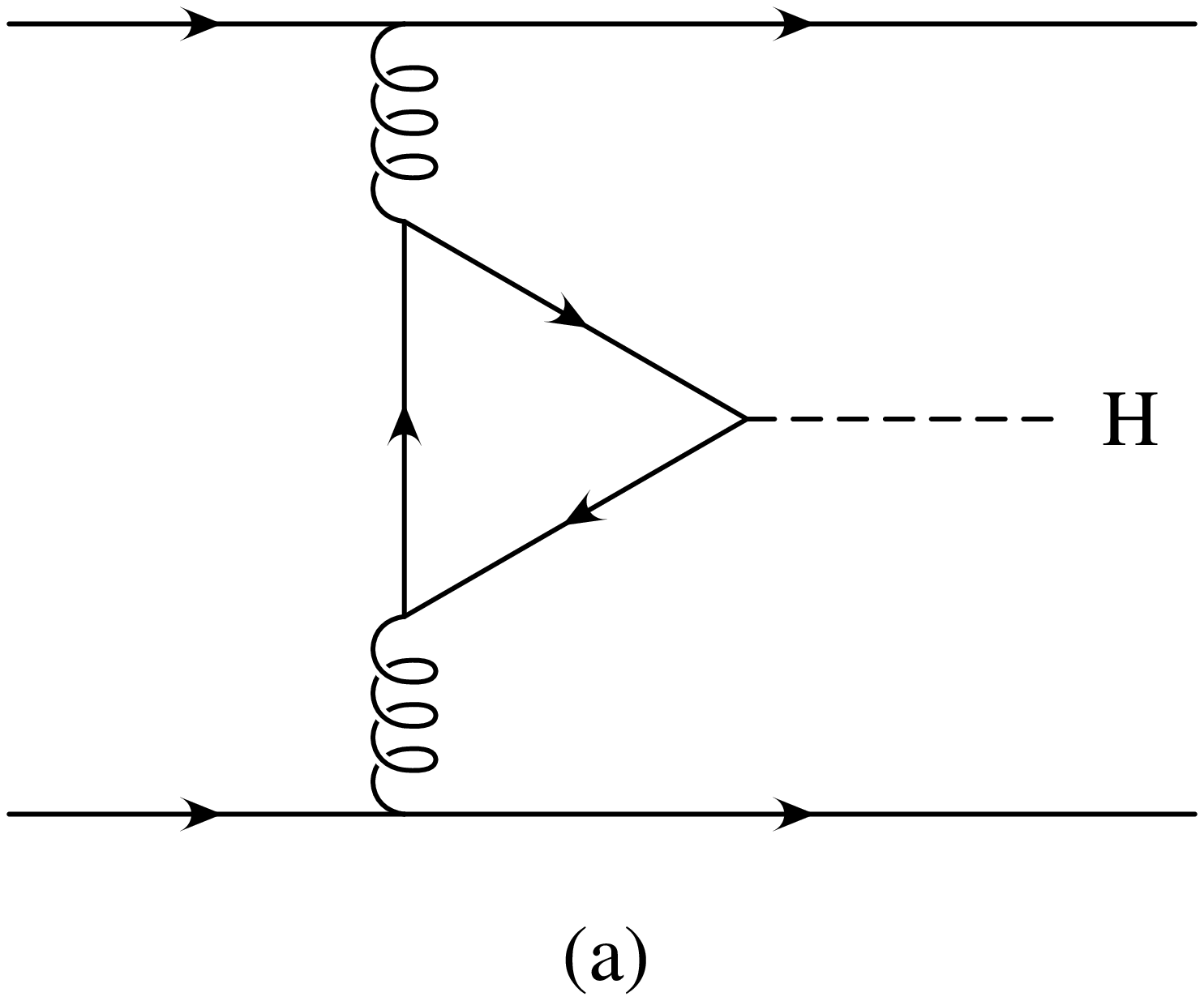,width=0.3\textwidth,clip=} \ \  
\epsfig{figure=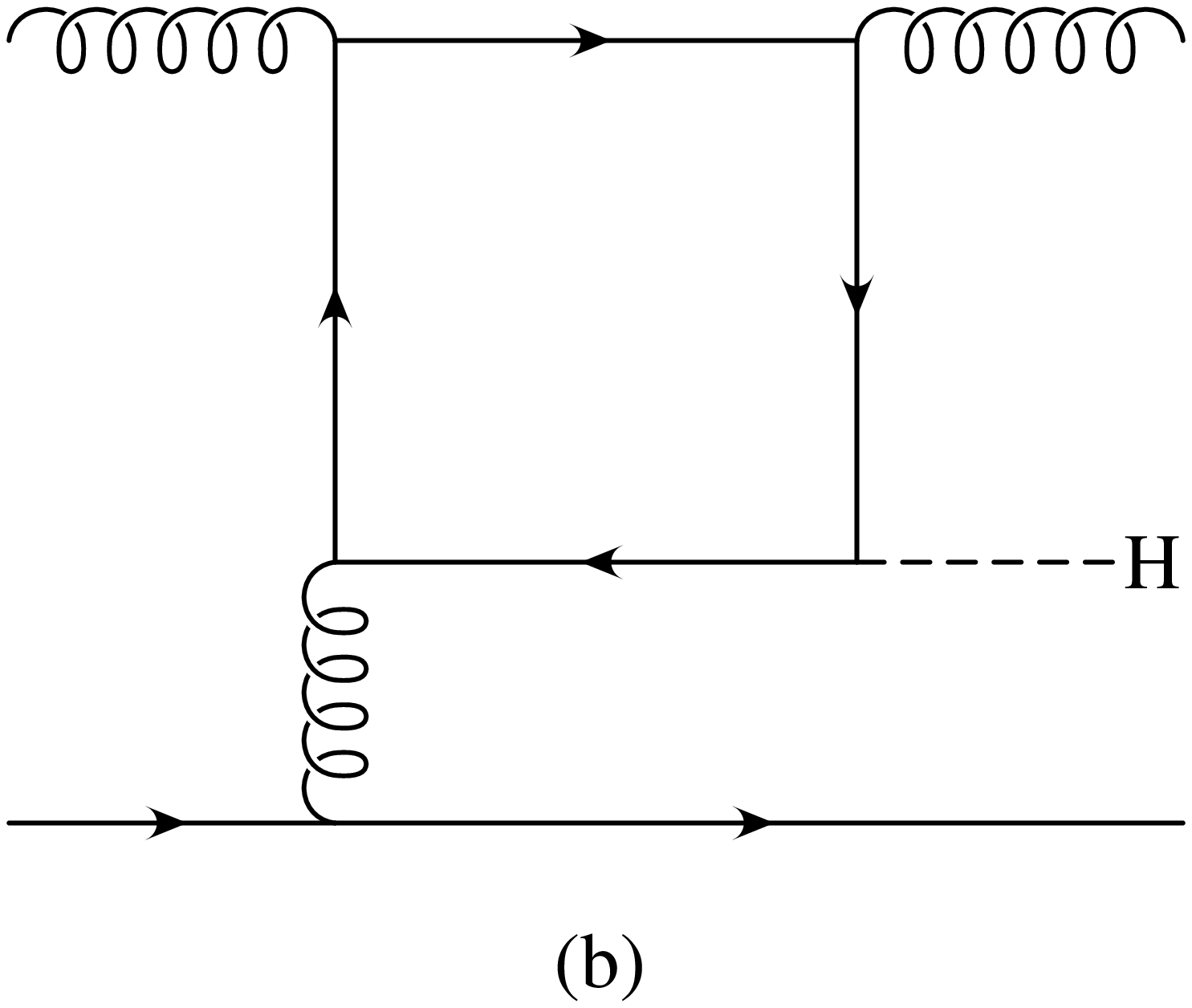,width=0.3\textwidth,clip=} \ \  
\epsfig{figure=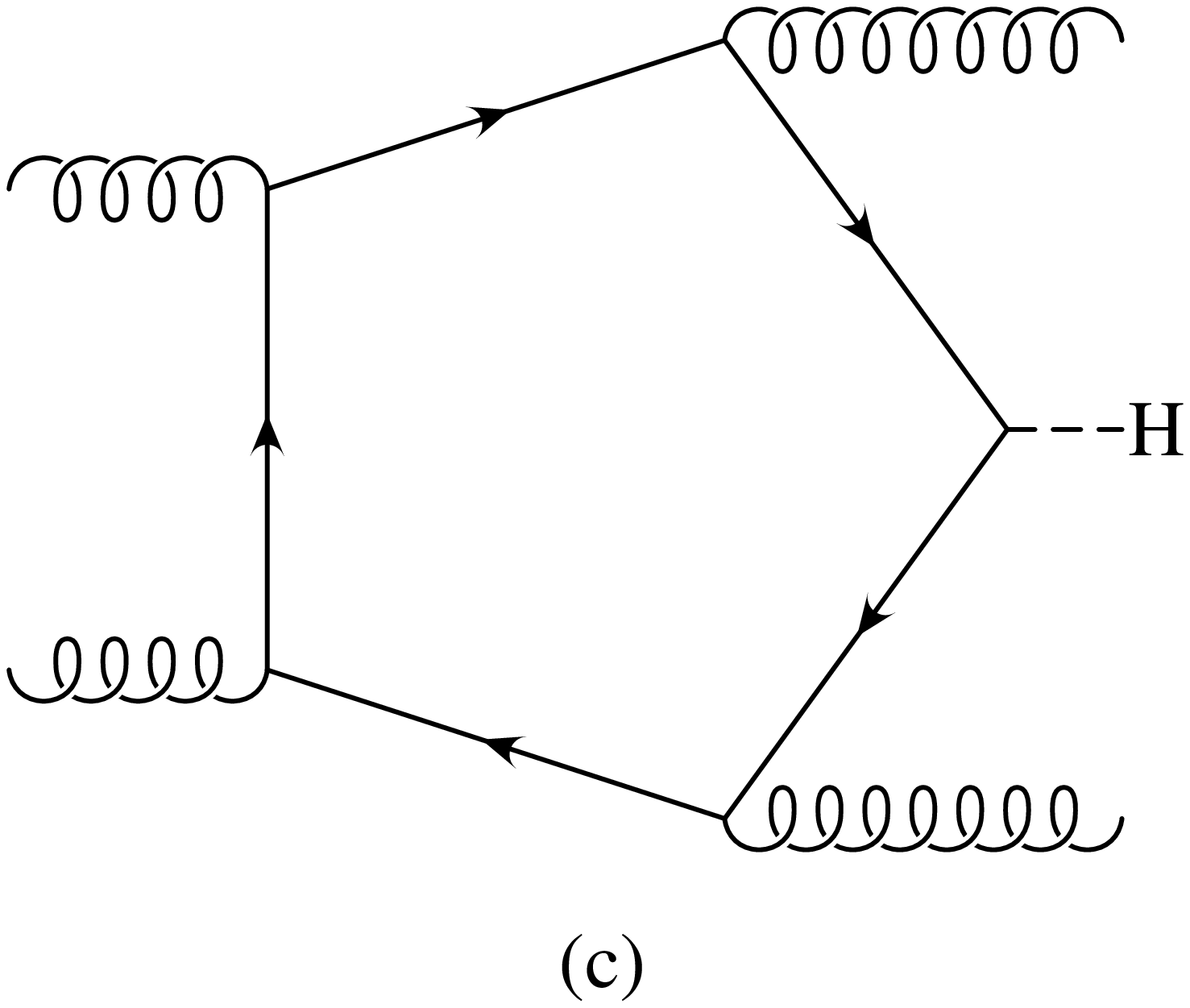,width=0.3\textwidth,clip=} \ \  
} 
\ccaption{} 
{ \label{fig:feyn} Examples of Feynman graphs contributing to $H+2$~jet 
production via gluon fusion. } 
\end{figure}

The production of a Higgs boson in association with two jets, at order
$\as^4$, can proceed via the subprocesses 
\bq 
\label{eq:processes} 
qq\,\to \,qqH\;,\qquad qQ\,\to \,qQH\;,\qquad qg \,\to \,qgH\;, \qquad
gg\,\to \,ggH\;,  
\eq 
and all crossing-related processes. Here the first two entries denote
scattering of identical and non-identical quark flavors.
In Fig.~\ref{fig:feyn} we have collected a few representative Feynman 
diagrams which contribute to subprocesses with four, two and zero external
quarks.  In our calculation, the top quark is treated as massive, but we  
neglect all other quark masses, so that the  
Higgs boson only couples via a top-quark loop. Typically we have a 
$ggH$ coupling through a triangle loop (Fig.~\ref{fig:feyn}~(a)), a $gggH$ 
coupling mediated by a box loop (Fig.~\ref{fig:feyn}~(b)) and a $ggggH$ 
coupling which is induced through a pentagon loop (Fig.~\ref{fig:feyn}~(c)). 
The number and type of Feynman diagrams can be easily built from the
simpler dijet QCD production processes at leading order. One needs to 
insert the Higgs-gluon ``vertices'' into the tree-level diagrams for 
$2\,\to\,2$ QCD parton scattering in all possible ways. 

In the following counting, we exploit 
Furry's theorem, i.e.\ we are counting as one the two charge-conjugation  
related diagrams where the loop momentum is  
running clockwise and counter-clockwise. This halves the number of diagrams.  
In addition, the crossed processes are not listed as extra diagrams, but are 
included in the final results. Three distinct classes of processes need to 
be considered.
 
\begin{enumerate} 
 \item {\boldmath $qq \,\to\, qq H$ and \boldmath $qQ \,\to\, qQ H$} 
 There are  only 2 diagrams obtained from the insertion of a triangle 
 loop into the tree-level diagrams for $ qq \,\to\, qq$. 
 One of them is depicted in Fig.~\ref{fig:feyn}~(a), while the other is 
 obtained by interchanging the two identical final quarks. 
 In the case of $qQ \,\to\, qQ H$, where $Q$ is a different flavor, there is 
 only one diagram, i.e.\ Fig.~\ref{fig:feyn}~(a). 
 
 \item {\boldmath $qg \,\to\, qg H$} At tree level, there are 3 diagrams 
 contributing to the process $qg \,\to\, qg$: one with a three-gluon 
 vertex and  two Compton-like ones. 
 Inserting a triangle loop into every gluon line, we have a total of 7 
 different diagrams. 
 In addition, we can insert a box loop into the diagram with the 
 three-gluon vertex, in 3 different ways: the $3!$ permutations of the  
 3 gluons are reduced to 3 graphs by using Furry's theorem.  
 In total we have 10 different diagrams for the $qg \,\to\, qg H$ 
 scattering amplitude. 
 
 \item {\boldmath $gg \,\to\, gg H$} Four diagrams contribute to 
 the tree-level scattering process $gg\,\to\, gg$: a four-gluon vertex 
 diagram and 3 diagrams with two three-gluon vertices each. 
 Inserting a triangle loop in any of the gluonic legs gives rise to 19 
 different diagrams. 
 The insertion of the box loop in the 3 diagrams with three-gluon vertices 
 yields another 18 diagrams. 
 Finally, there are 12 pentagon diagrams (corresponding to $4!$ permutations  
 of the external gluons, divided by 2, according to Furry's theorem). 
\end{enumerate} 
 
The amplitudes for these processes are ultraviolet and infrared finite in
$D=4$ dimensions. Nevertheless, we kept $D$ arbitrary in several parts of our
computation because some functions are divergent in $\ep=(4-D)/2$ 
at intermediate steps. 
Obviously, these divergences cancel when the intermediate expressions are
combined to give final amplitudes. An example of this behavior
is given by Eqs.~(\ref{eq:fl}) and~(\ref{eq:ft}), where the divergent part of
the $B_0$ functions cancels among the different contributions to the triangle
graphs. 

Given the large number of contributing Feynman graphs, it is most convenient 
to give analytic results for the scattering amplitudes for fixed 
polarizations of the external quarks and gluons. These amplitudes are then 
evaluated numerically, instead of using trace techniques to express 
polarization averaged squares of amplitudes in terms of relativistic 
invariants. We proceed to derive explicit expressions for these amplitudes.

\section{Notation and matrix elements} 
\label{sec:amp}

Within the SM, the effective interaction of the Higgs boson with gluons is 
dominated by top-quark loops because the top Yukawa coupling, $h_t=\m/v$ 
with $v=246.22$~GeV, is much larger than the $Hb\bar b$ coupling. In 
the following we only consider top-loop contributions. All the $H+4$~parton  
amplitudes, at lowest order, are then proportional to $h_t\,g_s^4$, where 
$g_s=\sqrt{4\pi\alpha_s}$ is the SU(3) coupling strength. It is convenient 
to absorb these coupling constants into an overall factor 
\bq 
F=h_t\;{g_s^4\over 16\pi^2}\; 4\,\m = 4\,{\m^2\over v}\,\alpha_s^2\;, 
\eq 
where we have anticipated the loop factor $1/16\pi^2$ and the emergence of an
explicit factor, $4\,\m$, from all top-quark loops, which results from the
compensation of the chirality flip, induced by the insertion of a single
scalar $Ht\bar t$ vertex.
 

\subsection{\boldmath $qQ\,\to\,qQH$ and  $qq\,\to\,qqH$} 

The simplest contribution to $H+2$~jet production is provided by the 
$qQ\,\to\, qQH$ process depicted in Fig.~\ref{fig:feyn}(a). Other four-quark 
amplitudes are obtained by crossing. We neglect external fermion masses 
and use the formalism and the notation of Ref.~\cite{HZ} for the spinor  
algebra. For a subprocess like 
\bq 
q(\overline p_1,i_1) + \overline Q(\overline p_4,i_4) \,\to\,  
q(\overline p_2,i_2) + \overline Q(\overline p_3,i_3) + H(P)\,,
\eq 
with
\beq
\overline p_1^2=\overline p_2^2=\overline p_3^2=\overline p_4^2=0,
\qquad P^2=m^2_H,
\eeq
each external (anti-)fermion is described by a two-component Weyl-spinor of  
chirality $\tau=\sigma_i=S_i\overline\sigma_i$,
\bq 
\psi(\overline p_i,\overline\sigma_i)_\tau =  
S_i\sqrt{2\,\overline p_i^0}\, \de_{\sigma_i\tau}
\chi_{\sigma_i}(\overline p_i)\;. 
\eq 
Here $\overline p_i$, $\overline \sigma_i$ and $i_i$ 
denote the physical momentum, the  
helicity and the color index of the quark or anti-quark,
and the sign factor $S_i$ allows   
for an easy switch between fermions ($S_i=+1$) and anti-fermions ($S_i=-1$).  
The quark-gluon vertices of Fig.~\ref{fig:feyn}(a), including the attached 
gluon propagators, are captured via the effective quark currents 
\bq 
\label{eq:J21} 
J_{21}^\mu = \delta_{\sigma_2\sigma_1}\; 
\chi^\dagger_{\sigma_2}(\overline p_2)(\sigma^\mu)_\tau  
        \chi_{\sigma_1}(\overline p_1) {1\over (p_1-p_2)^2} 
    = \delta_{\sigma_2\sigma_1}\; 
       \left<2\right| (\sigma^\mu)_\tau \left|1\right> {1\over (p_1-p_2)^2} 
\eq 
and 
\bq 
J_{43}^\mu = \delta_{\sigma_4\sigma_3}\; 
       \left<4\right| (\sigma^\mu)_\tau \left|3\right> {1\over (p_3-p_4)^2}
\;. 
\eq 
Here we have used helicity conservation via the assignments  
$\tau=\sigma_1=\sigma_2$ and $\tau=\sigma_3=\sigma_4$, respectively, and the  
sign factors for the fermions provide an easy connection between the gluon 
momenta $q_1=p_2-p_1$ and $q_2=p_4-p_3$ going out of the top-quark triangle 
and the physical quark momenta $\overline p_i = S_i p_i$. Finally we have  
used the shorthand notation  
\ba 
\left|1\right> &=& \chi_{\sigma_1}(\overline p_1) \;,\nonumber \\ 
\left<2\right| &=& \chi^\dagger_{\sigma_2}(\overline p_2) \;,\nonumber \\ 
\left|3\right> &=& \chi_{\sigma_3}(\overline p_3) \;,\nonumber \\ 
\left<4\right| &=& \chi^\dagger_{\sigma_4}(\overline p_4)\;, 
\ea 
and $(\sigma^\mu)_\pm = (1,\pm${\boldmath$\sigma$}) is the reduction of  
Dirac matrixes $\gamma^\mu$ into the two-component Weyl basis.
Since the quark currents $J_{12}^\mu$ and $J_{43}^\mu$ are conserved, we have
\beq
J_{12}^\mu  \(p_1-p_2\)_\mu = 0, \qquad J_{34}^\mu \(p_3-p_4\)_\mu = 0\;.
\eeq
The Weyl spinors and the currents $J_{21}$ and $J_{43}$ are easily evaluated  
numerically~\cite{HZ}. 
The scattering amplitude for different flavors on the 
two quark lines is then given by 
\bq 
{\cal A}^{qQ} =  F^{qQ} J_{21}^\mu J_{43}^\nu T_{\mu\nu}(q_1,q_2) 
 \, t^a_{i_2i_1}\, t^a_{i_4i_3} = 
{\cal A}_{2143}^{qQ}  \, t^a_{i_2i_1}\, t^a_{i_4i_3} \;, 
\eq 
where the overall factor 
\bq 
 F^{qQ} = S_1\,S_2\,S_3\,S_4\; 
4\sqrt{\overline p_1^0\,\overline p_2^0\,\overline p_3^0\,\overline p_4^0}\;F 
\eq 
includes the normalization factors of external quark spinors. The  
$t^a_{ij}={\lambda^a_{ij}/2}$ are the color generators in the 
fundamental representation of SU$(N)$, $N=3$. 
As detailed in Appendix~\ref{app:triangle}, the tensor $T_{\mu\nu}(p,q)$ can
be written as 
\beqn
T^{\mu\nu}(p,q) &=& F_T\(p^2,q^2,(p+q)^2\) \,  
\( p\cdot q\, g^{\mu\nu} - p^\nu \,q^\mu \)  \nonumber \\ 
& +& F_L\(p^2,q^2,(p+q)^2\) \,
\( q^2 p^2 g^{\mu\nu} - p^2 \,q^\mu \, q^\nu -  
q^2\, p^\mu \,p^\nu + p\cdot q \, p^\mu q^\nu \) \;.
\eeqn
Analytic expressions for the scalar form factors $F_T$ and $F_L$ are given
in Eqs.~(\ref{eq:fl}) and~(\ref{eq:ft}).
 
The scattering amplitude for two identical quarks is obtained from the  
result above by including Pauli interference, which results from interchanging 
quarks 2 and 4, 
\bq 
{\cal A}^{qq} = {\cal A}_{2143}^{qq}  \, t^a_{i_2i_1}\, t^a_{i_4i_3} 
               -{\cal A}_{4123}^{qq}  \, t^a_{i_4i_1}\, t^a_{i_2i_3}\; . 
\eq 
The squared amplitude, summed over initial- and final-particle color, 
becomes  
\bq 
\label{eq:A_qq} 
\sum_{\rm col}|{\cal A}^{qq}|^2 =  
\(|{\cal A}_{2143}|^2 + |{\cal A}_{4123}|^2 \)\frac{N^2-1}{4} +  
2 \Re\({\cal A}_{2143} {\cal A}_{4123}^*\) \frac{N^2-1}{4N}. 
\eq 
The squared amplitude  for the $qQ \,\to\, qQ H$ process can be read from  
Eq.~(\ref{eq:A_qq}) by putting ${\cal A}_{4123}=0$.

\subsection{\boldmath $qg\,\to\,qgH$} 
\label{sec:qqgg}
\begin{figure}[htb] 
\centerline{ 
\epsfig{figure=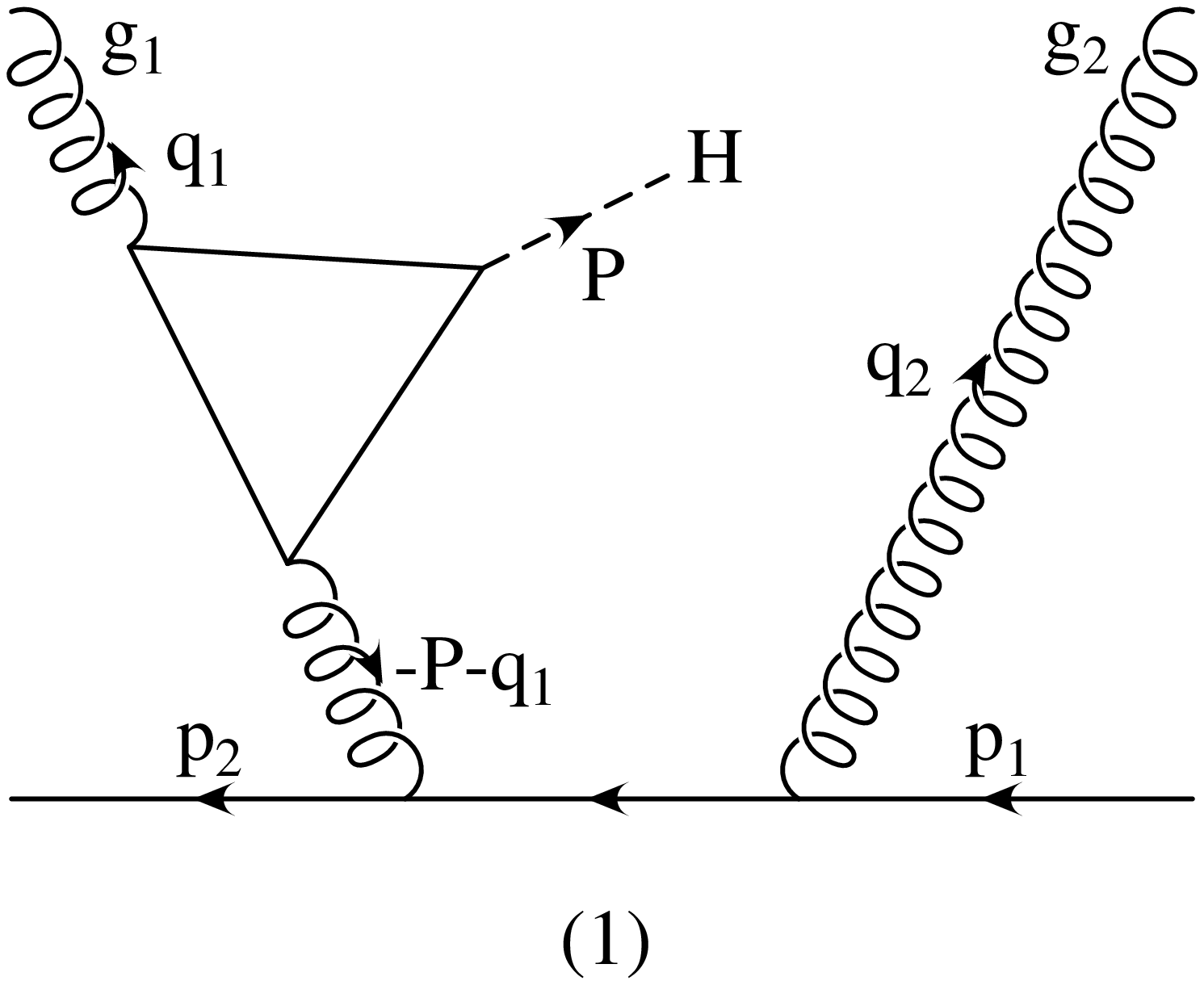,width=\largfig,clip=} \ \ \ \ 
\epsfig{figure=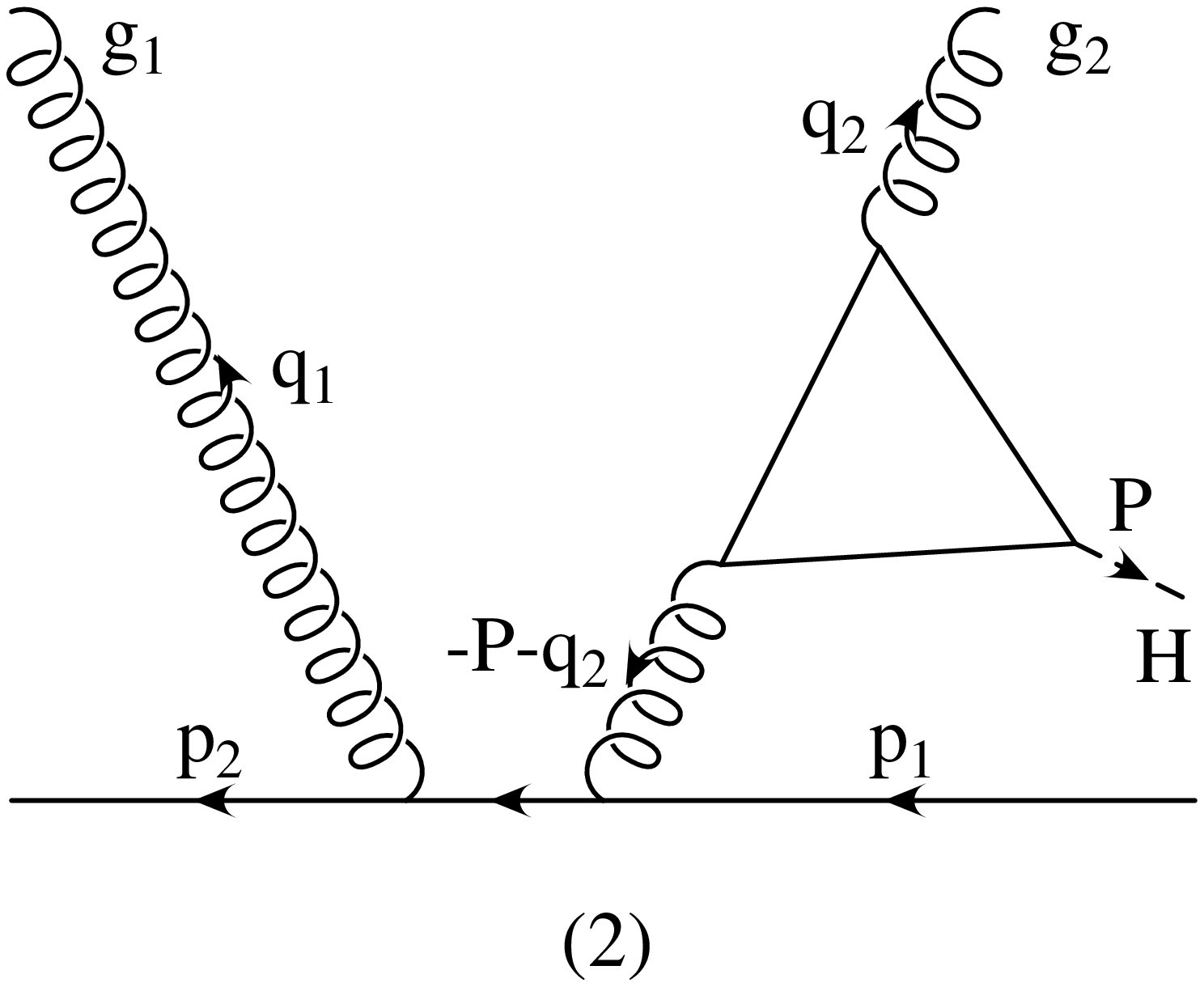,width=\largfig,clip=} \ \  \ \ 
\epsfig{figure=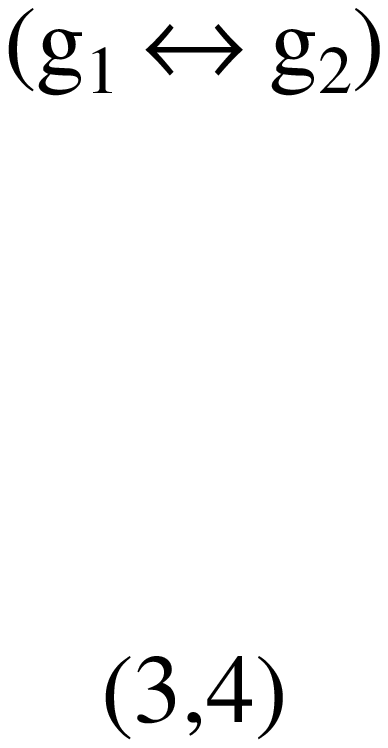,width=\largfig,clip=}
}
\vspace{0.5cm}
\centerline{ 
\epsfig{figure=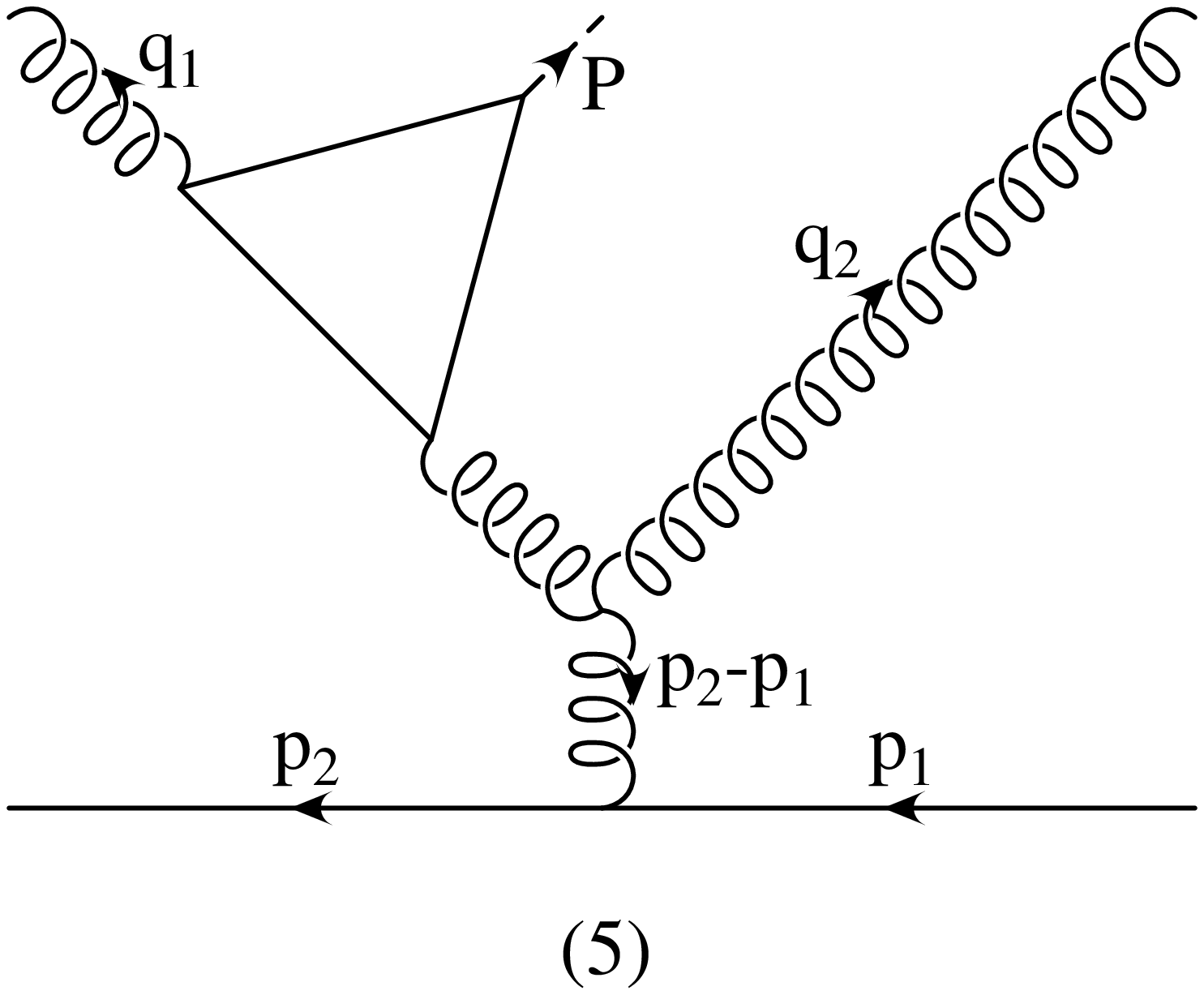,width=\largfig,clip=} \ \ \ \
\epsfig{figure=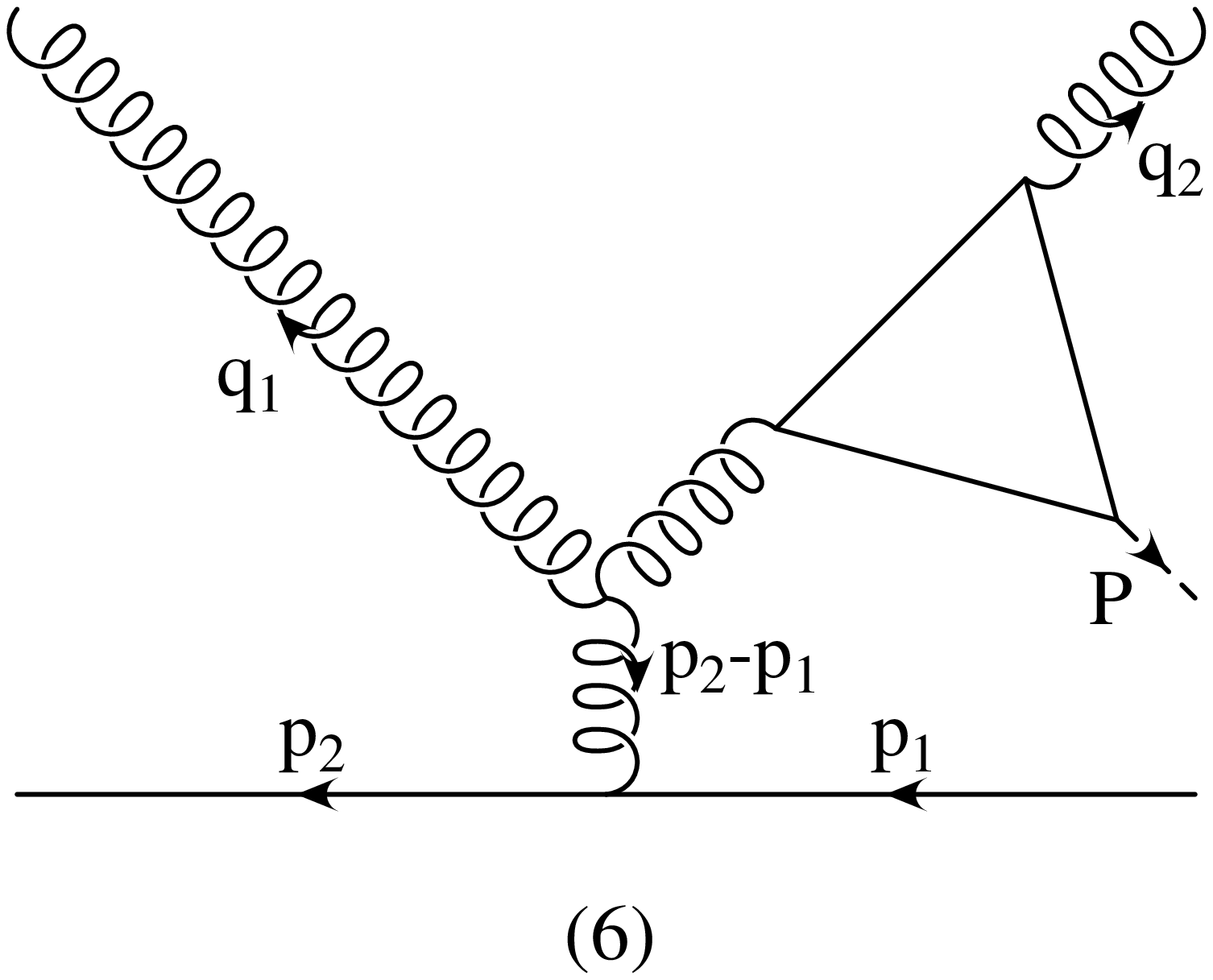,width=\largfig,clip=} \ \ \ \
\epsfig{figure=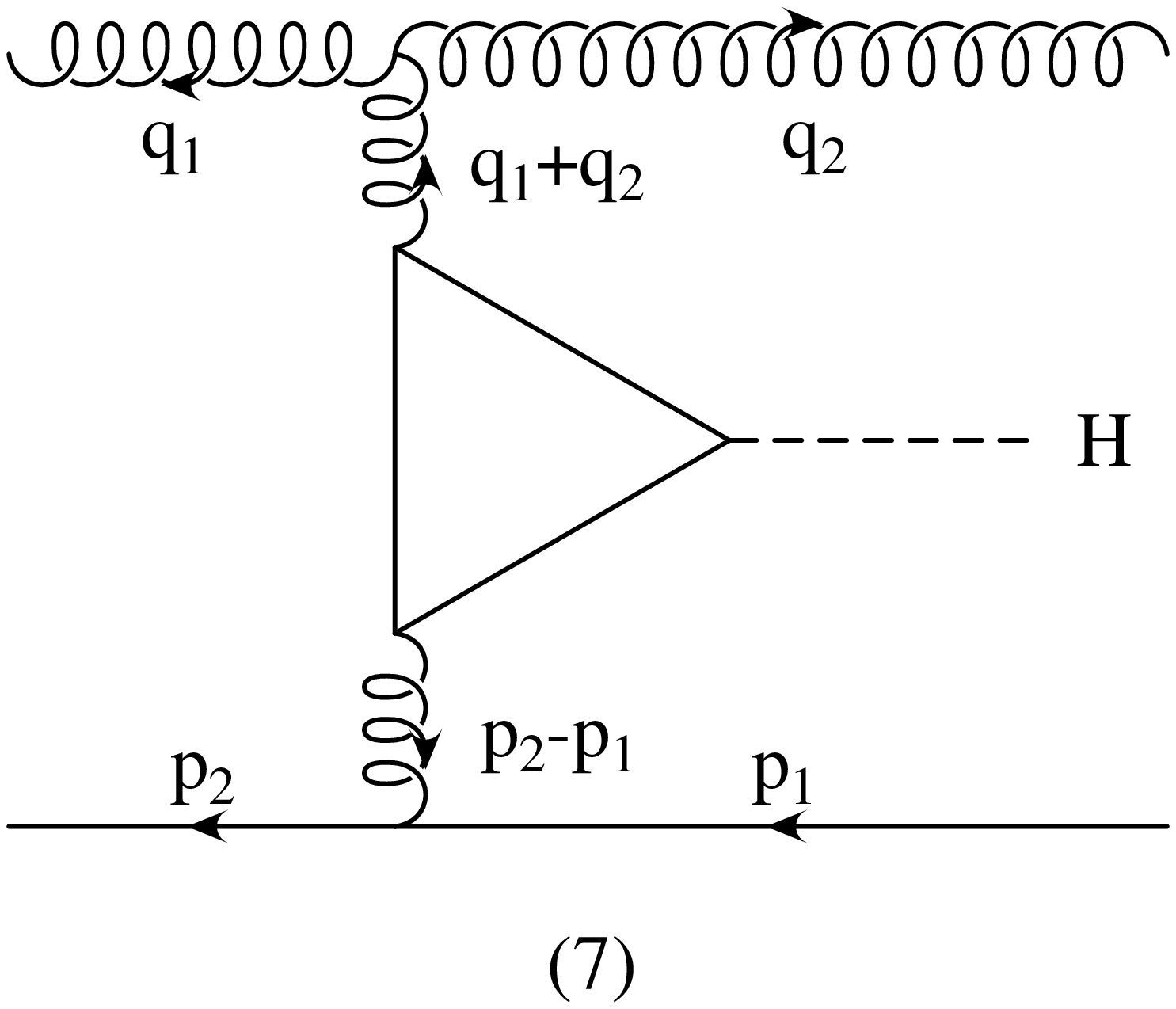,width=\largfig,clip=} 
}
\vspace{0.5cm}
\centerline{ 
\epsfig{figure=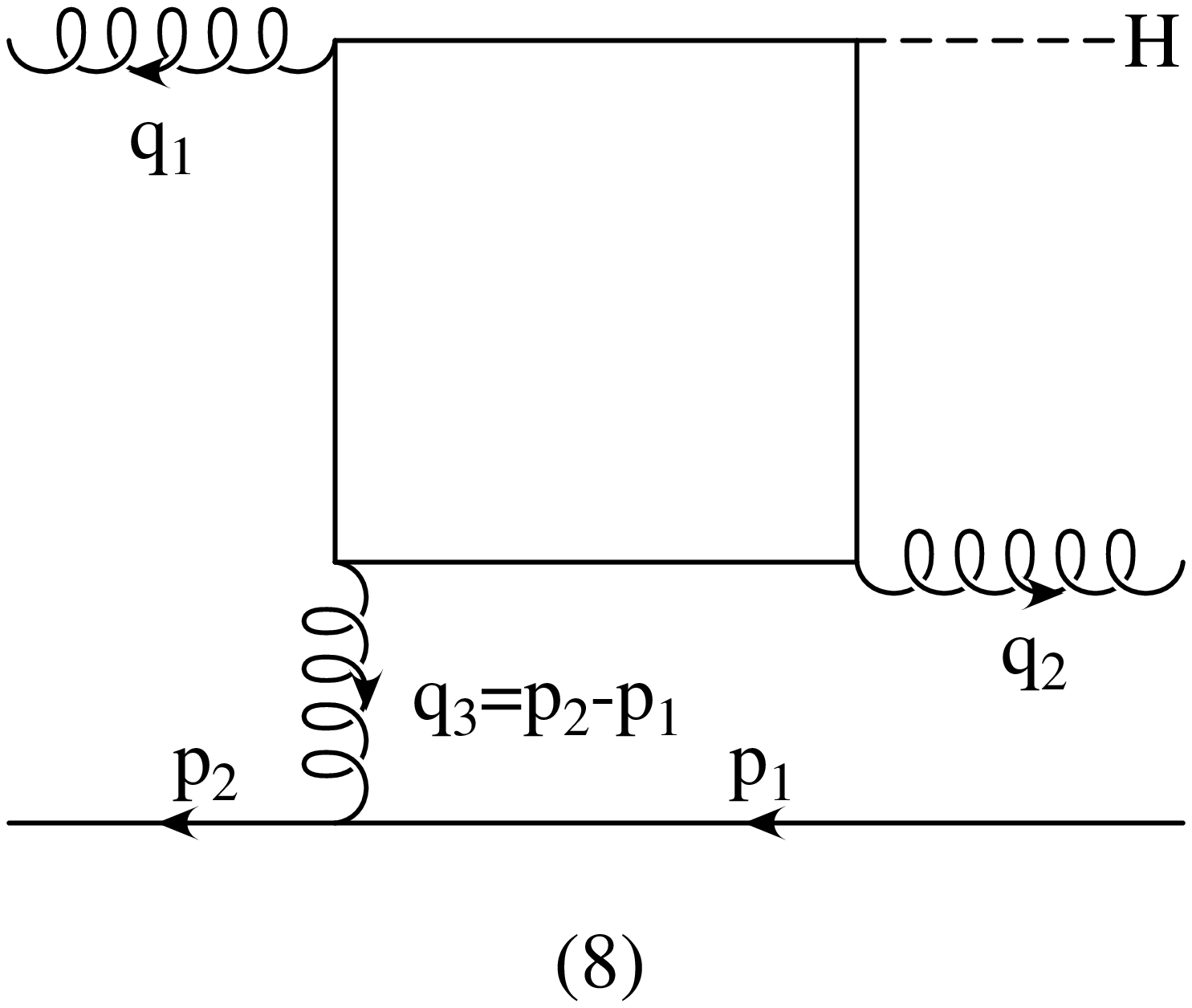,width=\largfig,clip=} \ \  \ \
\epsfig{figure=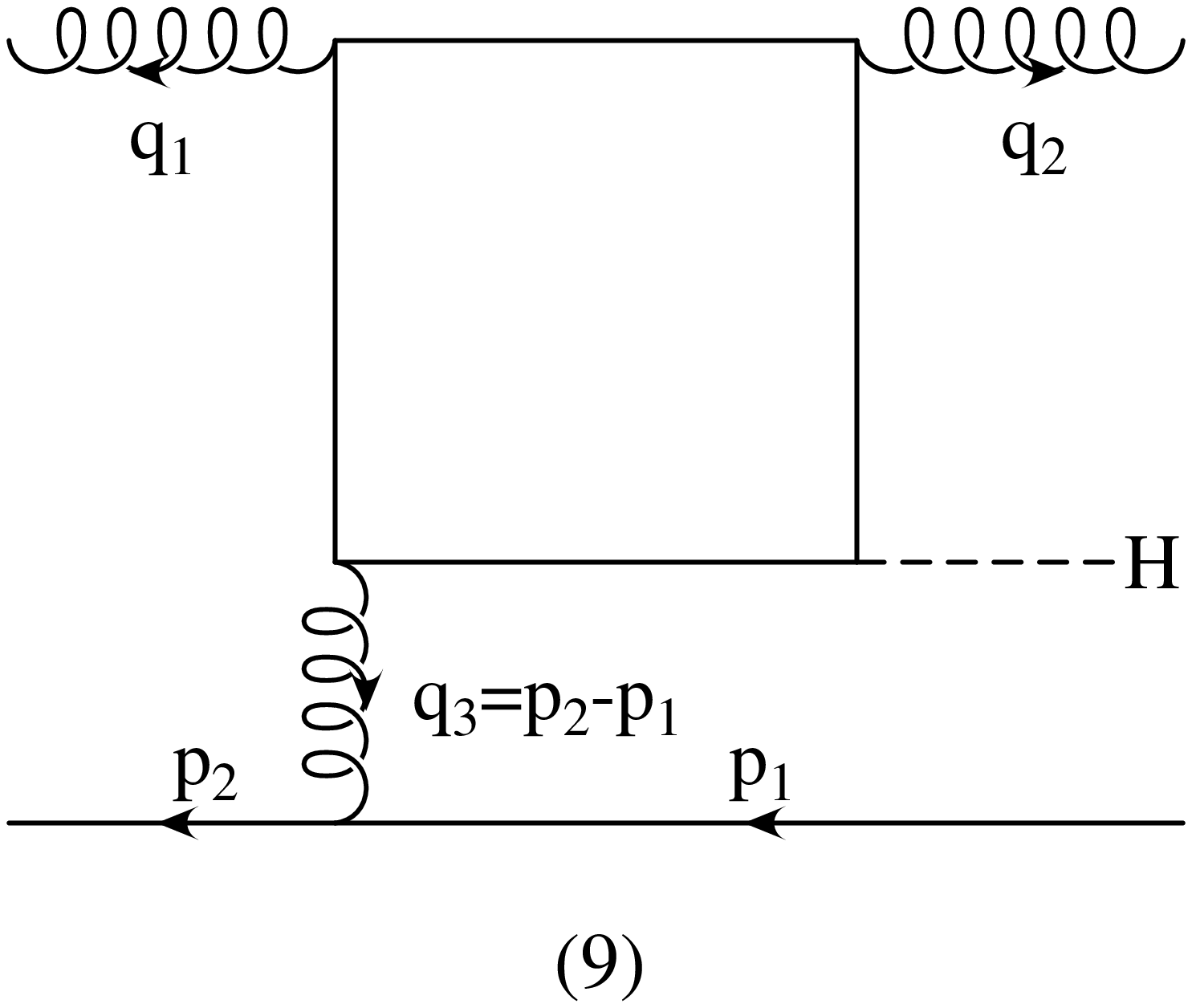,width=\largfig,clip=} \ \  \ \
\epsfig{figure=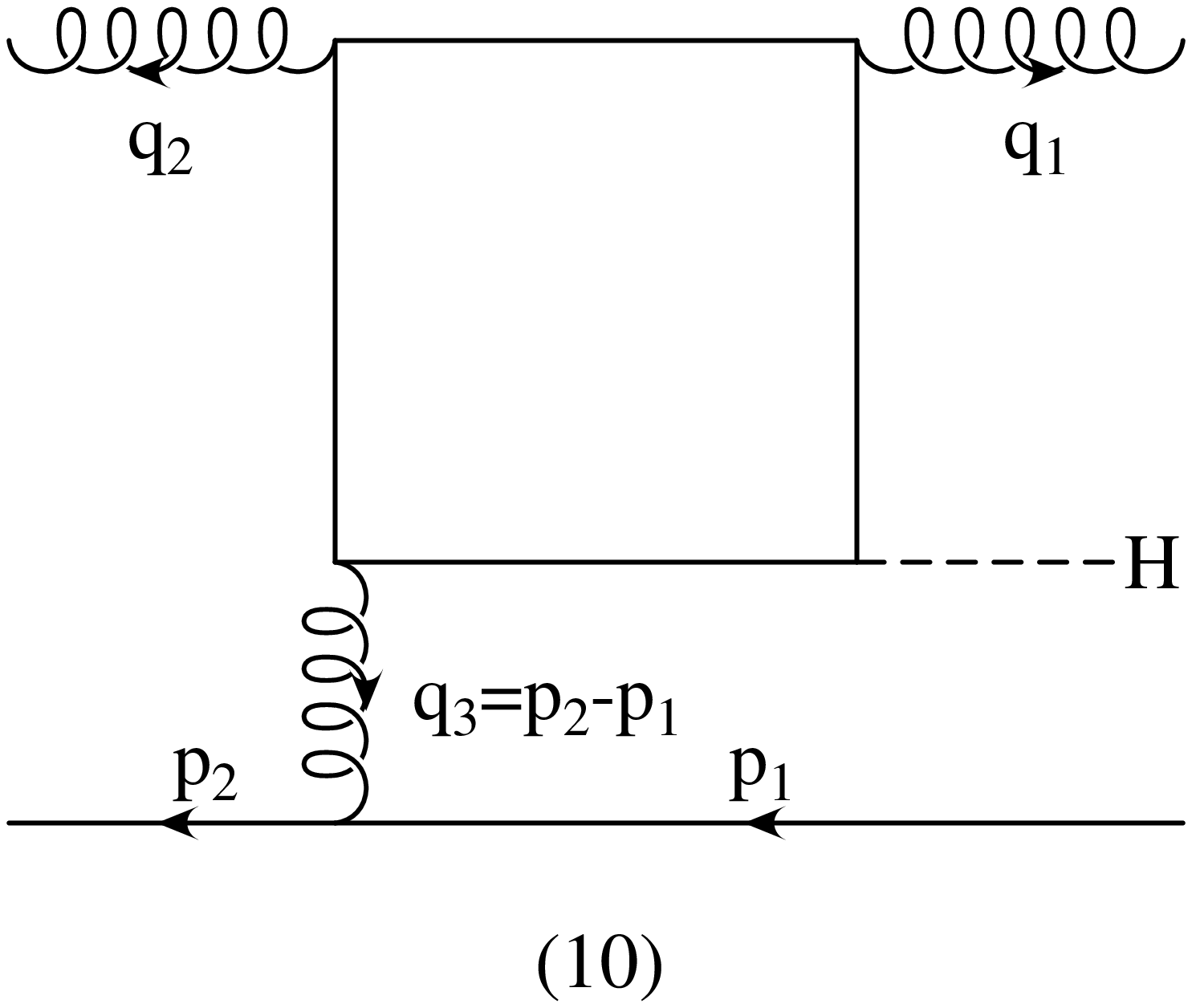,width=\largfig,clip=} 
}
\ccaption{} 
{ \label{fig:feyn_gq} 
Feynman graphs contributing to the process $qg\to qgH$. Graphs (3) and (4)
are the same as (1) and (2), but with gluon labels interchanged. No 
distinction is made between the two 
orientations of the fermion arrow on the top-quark loop, 
because they are related by Furry's theorem.}
\end{figure} 
 
Twenty distinct Feynman graphs contribute to the process 
\bq 
\label{eq:proc_gq} 
g(\overline q_1,a_1)+q(\overline p_1,i_1)\,\to  \,
g(\overline q_2,a_2)+q(\overline p_2,i_2)+H(P)\;,
\eq 
and crossing related processes (the color index of the external gluons is
indicated  with $a_i$).
However, pairs with opposite directions of 
the top-quark fermion arrow in the loop are related by charge conjugation 
and will not be counted separately in the following (see
Appendixes~\ref{app:triangle} and~\ref{app:boxes}).  

The resulting ten Feynman graphs are depicted in Fig.~\ref{fig:feyn_gq}. 
Following Ref.~\cite{HZ}, all gluon momenta are treated as outgoing. 
For the specific process of Eq.~(\ref{eq:proc_gq}) we then set  
$q_1=-\overline q_1$ and $q_2=+\overline q_2$, with
\beq
\overline p_1^2=\overline p_2^2=\overline q_1^2=\overline q_2^2=0,
\qquad P^2=m^2_H.
\eeq
Feynman graphs with a triangle insertion in an external gluon line, 
i.e.\ with one light-like gluon attached to the top-quark triangle, receive 
contributions from a single form factor, $F_T$, only (see
Eq.~(\ref{eq:ep_T_L})).
These simplifications are best captured by replacing the polarization 
vectors $\eps_i^\mu = \eps_i^\mu(\overline q_i)$ with the
effective  polarization vectors  
\bq 
e^\mu_{iH} = F_T\(0,\(P+q_i\)^2,P^2\) {1\over \(P+q_i\)^2} 
\( q_i^\mu \, \eps_i\cdot P - \eps_i^\mu \,q_i\cdot P \)\;, 
\eq 
for gluons $i=1,2$, with $F_T$ given in Eq.~(\ref{eq:ft}).  
 
External quark lines are handled as for the $qQ\to qQH$ amplitudes.  
Feynman graphs 5 through 10 are proportional to the quark current  
$J_{21}^\mu$ defined in Eq.~(\ref{eq:J21}). Spinor normalization factors  
are absorbed into the overall factor 
\bq 
F^{qg} = -S_1S_2\, 
2\sqrt{\overline p_1^0\,\overline p_2^0}\;F\;\delta_{\sigma_1\sigma_2}\;. 
\eq 
Using the shorthand notation~\cite{HZ} 
\beqn 
\left<2,q_i\right| &=& \chi^\dagger_{\sigma_2}(\overline p_2) 
(\sla\eps_i)_{\sigma_2}(\sla p_2+\sla q_i)_{-\sigma_2} 
{1\over (p_2+q_i)^2} \;,
\\ 
\left|q_i,1\right> &=& (\sla p_1-\sla q_i)_{-\sigma_1}(\sla\eps_i)_{\sigma_1} 
 \chi_{\sigma_1}(\overline p_1) {1\over (p_1-q_i)^2} \;,
\eeqn 
for the 2-component Weyl spinors describing emission of a gluon next to  
an external quark, we arrive at a compact notation for the contributions 
to the $qg\,\to\, qgH$ scattering amplitude 
\beqn 
\label{eq:amp_qg} 
{\cal A}^{qg} \!\!\!\! &=& \! \!\!{\cal A}^{qg}_{\mu_1\mu_2}\, \eps_1^{\mu_1}\,
\eps_2^{\mu_2}  \nonumber \\ 
 &=&  \!\!\!\! F^{qg}\Bigg\{ \!\! 
\(t^{a_1}t^{a_2}\)_{i_2i_1} \Big[  
\left<2\right|(\sla e_{1H})_{\sigma_1}\left|q_2,1\right> + 
\left<2,q_1\right|(\sla e_{2H})_{\sigma_1}\left|1\right> \Big] +  
\nonumber \\ &&  \phantom{\!\!\!\! F^{qg}\Bigg\{ \!\!} 
 \(t^{a_2}t^{a_1}\)_{i_2i_1}\Big[
\left<2\right|(\sla e_{2H})_{\sigma_1}\left|q_1,1\right> + 
\left<2,q_2\right|(\sla e_{1H})_{\sigma_1}\left|1\right> \Big] +  
\nonumber \\ && \phantom{\!\!\!\! F^{qg}\Bigg\{ \!\!}
\lq t^{a_1},t^{a_2}\rq_{i_2i_1}\Biggl[\mbox{}  
 +2\Big( J_{21}\cdot q_2\,  e_{1H}\cdot \eps_2 - J_{21}\cdot
e_{1H} \, \(p_2-p_1\)\cdot\eps_2  
   - J_{21}\cdot\eps_2 \, q_2\cdot e_{1H}\Big)  
\nonumber \\ &&  \phantom{\!\!\!\! F^{qg}\Bigg\{ \!\!
\lq t^{a_1},t^{a_2}\rq_{i_2i_1} \biggl[ }
-2\Big( J_{21}\cdot q_1 \, e_{2H}\cdot \eps_1 - J_{21}\cdot e_{2H}\, 
\(p_2-p_1\)\cdot\eps_1 - J_{21}\cdot\eps_1 \, q_1\cdot e_{2H}\Big)  
\nonumber \\ && \phantom{\!\!\!\! F^{qg}\Bigg\{ \!\!
\lq t^{a_1},t^{a_2}\rq_{i_2i_1} \Biggl[ }
-\lq \( \frac{\(q_1+q_2\)\cdot \(p_1-p_2\)}{(q_1+q_2)^2} F_T - 
       \(p_1-p_2\)^2 F_L \) J_{21}^\mu 
+ \frac{J_{21}\cdot P}{(q_1+q_2)^2} F_T\, P^\mu \rq
\nonumber \\ && \phantom{\!\!\!\! F^{qg}\Bigg\{ \!\!
\lq t^{a_1},t^{a_2}\rq_{i_2i_1} \biggl[ -\Bigg[}
\times \Big[ \eps_1\cdot\eps_2 \, \(q_2-q_1\)_\mu
-2\,q_2\cdot\eps_1\,\eps_{2\mu}  
      +2\,q_1\cdot\eps_2\, \eps_{1\mu}\Big]  
\nonumber \\ && \phantom{\!\!\!\! F^{qg}\Bigg\{ \!\!
\lq t^{a_1},t^{a_2}\rq_{i_2i_1} \biggl[ }
- B_{\mu_1\mu_2\mu_3}\,\eps_1^{\mu_1}\,\eps_2^{\mu_2}\, J_{21}^{\mu_3} \Biggr] 
\Bigg\} 
\eeqn 
where $F_T = F_T\(\(q_1+q_2\)^2,\(p_1-p_2\)^2,P^2\)$,  
and analogously for $F_L$. 
 
The contributions of the box diagrams enter in the last line of 
Eq.~(\ref{eq:amp_qg}) via the tensor 
$B_{\mu_1\mu_2\mu_3}=B_{\mu_1\mu_2\mu_3}(q_1,q_2,q_3)$ (with $q_3=p_2-p_1$). 
Gauge invariance and Bose symmetry of the 
gluons limit the relevant structure of these box contributions to just two
independent scalar functions, as we will now show.

It is easy to see that the $qg\,\to\, qgH$ amplitude ${\cal A}^{qg}$ is
invariant under the replacements $\eps_1^{\mu}\,\to\, \eps_1^{\mu} + \kappa_1
q_1^\mu$ and $\eps_2^{\mu}\,\to\,\eps_2^{\mu} + \kappa_2 q_2^\mu$, for
arbitrary constants $\kappa_i$.\footnote{These gauge invariance conditions
provided a stringent test of our numerical programs.}  By proper choice of
these constants, the polarization vectors of the two on-shell gluons can be
made orthogonal to both $q_1$ and $q_2$. This can be seen by
introducing a convenient basis of Minkowski space, composed of the vectors
$q_1^\mu$, $q_2^\mu$ and 
\beqn
x^\mu &=&  q_2\cdot q_3 \, q_1^\mu + q_1\cdot q_3\, q_2^\mu  
	- q_1\cdot q_2 \,  q_3^\mu \;,\\
y^\mu &=& \eps^{\mu\alpha\beta\rho}q_{1\alpha}\, q_{2\beta}\, q_{3\rho}\;.
\eeqn
The vector $y^\mu$ is orthogonal to all momenta occurring in the
boxes ($q_1, q_2$ and $q_3$) while 
$x^\mu$ is orthogonal to $q_1^\mu$, $q_2^\mu$ and $y^\mu$. More precisely
\beqn
x\cdot x &=& \detx \equiv  q_1\cdot q_2 
\Bigl[q_3^2 \; q_1\cdot q_2 -2\;q_1\cdot q_3\;q_2\cdot q_3\Bigr]\;,
\label{eq:detx}
\\
y\cdot y &=& \detx \;,\\
\label{eq:x_prods}
x\cdot q_1&=&0\;,\qquad x\cdot q_2=0\;,\qquad 
x\cdot q_3=-{\detx\over q_1\cdot q_2}\;, 
\\
\label{eq:y_prods}
y\cdot q_1&=&0\;,\qquad y\cdot q_2=0\;,\qquad 
y\cdot q_3=0\;, \qquad y\cdot x = 0\;,
\eeqn
where $\detx$ denotes the Gram determinant of the box, i.e.\ the determinant
of the $3\times 3$ matrix with elements $\({\cal
Q}_3\)_{ij}=-q_i\cdot q_j$.
Obviously $\detx$ is symmetric under interchange of the three gluon labels.
The non-symmetric form given in Eq.~(\ref{eq:detx}) pertains to our particular
situation where only $q_3$ may be off-shell ($q_1^2=q_2^2=0$).

In the following we take 
\beqn
\label{eq:pol_vec1}
\ep_1^\mu &=& \frac{x^\mu}{\sqrt{-\detx}}\,,\\
\label{eq:pol_vec2}
\ep_2^\mu &=& \frac{y^\mu}{\sqrt{-\detx}}\,,
\eeqn
as the two independent polarization vectors of each of the on-shell gluons.
With this choice, we 
eliminate the two $B_c$ contributions in the box tensor 
$B_{\mu_1\mu_2\mu_3}$ (see Eq.~(\ref{eq:boxtensor14}))
since they contain a factor $q_1^{\mu_2}$
or $q_2^{\mu_1}$, which vanishes upon contraction with the polarization
vectors. 

We can then write the squared element for $qg \,\to\, qgH$, summed over the
polarization vectors of the external gluons in the form
\bq
\sum_{\rm pol}|{\cal A}^{qg}|^2 =
\({1\over \detx}\)^2 \lq
|{\cal A}_{xx}^{qg}|^2+|{\cal A}_{xy}^{qg}|^2+|{\cal A}_{yx}^{qg}|^2+|{\cal
A}_{yy}^{qg}|^2 \rq \;,
\eq
where the shorthand ${\cal A}_{xy}^{qg}={\cal A}^{qg}_{\mu_1\mu_2}
\, x^{\mu_1}\, y^{\mu_2}$, etc. has been used. Expressions for the contracted
tensor integrals for the boxes that appear in Eq.~(\ref{eq:amp_qg}) are given
in Eqs.~(\ref{eq:B_yyu})--(\ref{eq:B_xxv}).

In addition, the color structure of the 
$qg\to qgH$ amplitude is given by (see Eq.~(\ref{eq:amp_qg})) 
\bq
{\cal A}^{qg} = (t^{a_1}t^{a_2})_{i_2i_1} {\cal A}^{qg}_{12} +
(t^{a_2}t^{a_1})_{i_2i_1} {\cal A}^{qg}_{21} \;,
\eq
so that the resulting color-summed squared amplitude takes the form
\beq 
\label{eq:colA_qg} 
\sum_{\rm col}|{\cal A}^{qg}|^2 =   
\(|{\cal A}^{qg}_{12}|^2 + |{\cal A}^{qg}_{21}|^2 \)\frac{\(N^2-1\)^2}{4N} - 
2 \Re\lq{\cal A}^{qg}_{12} \({\cal A}^{qg}_{21}\)^*\rq \frac{N^2-1}{4N}
\;. 
\eeq

\subsection{\boldmath $gg\,\to\,ggH$} 
\label{sec:gg_ggH}
For the process
\beq 
g(\overline q_1,a_1) + g(\overline q_2,a_2) \, \to \, 
g(\overline q_3,a_3) + g(\overline q_4,a_4) + H(P)\;,
\eeq
we introduce the outgoing momenta $q_i$, so that $q_1 = - \overline q_1$, 
$q_2 = - \overline q_2$, $q_3 =  +\overline q_3$ and $q_4 = + \overline q_4$,
where 
\beq
\overline q_1^2=\overline q_2^2=\overline q_3^2=\overline q_4^2=0,
\qquad P^2 = m_H^2,
\eeq 
and $a_i$ are the color indices in the adjoint representation carried by 
the gluons. 
  
Due to the large number of diagrams and the length of the
results, we are not 
going to write explicitly the expressions for the amplitude, but we describe 
in detail the procedure we follow. 
 
We used {\tt QGRAF}~\cite{QGRAF} to generate the 49 diagrams for this process. 
As detailed in Sec.~\ref{sec:calculation}, these diagrams are obtained by
the insertion of a triangle, a box or a pentagon loop into the tree-level
diagrams for $gg\,\to\, gg$ scattering, so that we can write the un-contracted
amplitude in the ``formal''
way 
\beq 
\label{eq:uncontract_gg} 
\({\cal A}^{gg}\)_{\al\bt\ga\de} \equiv  
    \sum_{i=1}^{19} c_i^T \, T^i_{\al\bt\ga\de} + 
     \sum_{i=1}^{18} c_i^B \, B^i_{\al\bt\ga\de} + 
     \sum_{i=1}^{12} c_i^P \, P^i_{\al\bt\ga\de}. 
\eeq 
where the tensor functions $T^i$, $B^i$ and $P^i$ are Feynman diagrams that
contain a triangle, a box and a pentagon fermionic loop, while $c_i^T, c_i^B$
and $c_i^P$ are the respective color factors.
This amplitude will be
contracted with the external polarization vectors of the gluons, $\ep_i$, to
give 
\beq 
\label{eq:amp_gg} 
  {\cal A}^{gg} = \ep_{1}^{\al}\, \ep_{2}^{\bt} \, \ep_{3}^{\ga} \, 
   \ep_{4}^{\de} \,   
   \({\cal A}^{gg}\)_{\al\bt\ga\de} \equiv 
  \({\cal A}^{gg}\)_{\ep_1\ep_2\ep_3\ep_4}.
\eeq 
We used {\tt Maple} to trace over the Dirac $\ga$ matrixes and to 
manipulate the expressions.  
Since the Higgs couples as a scalar to the massive top quark in the loop, the 
resulting trace in the numerator of the generic $n$-point function has at 
most $n-1$ loop-momentum factors. 
Using the tensor-reduction procedure described by Passarino and 
Veltman~\cite{PV}, we can express the triangle and box one-loop tensor 
integrals in terms of the external momenta $q_i^{\mu}$ and in terms of the 
scalar functions $C_{ij}$ ($i=1\ldots 2,\ j=1\ldots 4$) and $D_{ij}$  
($i=1\ldots 3,\ j=1\ldots 13$). 
For speed reasons, 
we preferred not to express directly the final 
amplitude in terms of scalar triangles and boxes (the $C_0$ and $D_0$ 
functions in Passarino-Veltman notation), in our Monte Carlo program. 
Expressions for the amplitude written in terms of the  $C_0$ and $D_0$
integrals  
are considerably larger than the result where we keep the $C_{ij}$ 
and $D_{ij}$ functions. 
 
In dealing with diagrams with a pentagon loop,  
we worked  
directly with the dot products in the numerator. In fact, the generic 
tensor pentagon appearing in $gg\,\to\,ggH$ scattering has the form
\beqn 
\label{eq:pent_tens_def} 
&&E(p_1,p_2,p_3,p_4)^{\{\al,\ \al\bt,\ \al\bt\ga,\ \al\bt\ga\de\}} = 
\nonumber \\  
&& {}\hspace{-5mm}\int \frac{d^Dk}{i\pi^{D/2}}  
\frac{\{k^\al,\ k^\al k^\bt, \ k^\al k^\bt k^\ga,\  k^\al k^\bt k^\ga k^\de \}} 
{[k^2-\m^2][(k+p_1)^2-\m^2][(k+p_{12})^2-\m^2] 
       [(k+p_{123})^2-\m^2][(k+p_{1234})^2-\m^2]},\nonumber\\ 
\eeqn 
where $p_{ij}=p_i+p_j$, and similar ones for $p_{ijl}$ and $p_{ijln}$, and
where the set of $p_i$, $\{p_1,p_2,p_3,p_4\}$, is one of the 24 permutations
of the external gluon momenta $\{q_1,q_2,q_3,q_4\}$. These permutations are
reduced to 12, once Furry's theorem is taken into account.  The generic
scalar five-point function, that is Eq.~(\ref{eq:pent_tens_def}) with a 1 in
the numerator, will be indicated with $E_0(p_1,p_2,p_3,p_4)$.

The tensor indices appearing in the numerator are always contracted with one
of the following:
\begin{enumerate} 
\item the metric tensor $g^{\mu\nu}$ 
\item one of the external momenta $q_{i}^{\mu}$  
\item one of the external polarization vector $\ep_{i}^{\mu}$. 
\end{enumerate} 
The procedure we have used in these three cases is the following. 
\begin{enumerate} 
\item Every time there is a $k^2=k^\al k^\bt g_{\al\bt}$ product in the 
numerator, we write it as  
\beq 
k^2 = \lq k^2-\m^2\rq + \m^2, 
\eeq 
and the first term is going to cancel the first propagator, giving rise to a 
four-point function, while the last one will multiply the rest of the tensor 
structure in the numerator, that now has been reduced by two powers of the 
loop momentum $k$.

\item We rewrite every scalar product of the type $(k \cdot q_i)$ in the 
numerator as a difference of two propagators, using the identity 
\beq 
\label{eq:k_dot_qi} 
k \cdot q_i = \frac{1}{2} \lg \lq \(k+p+q_i\)^2 -\m^2\rq - 
\lq \(k+p\)^2 -\m^2\rq -2 \,q_i\cdot p \rg,
\eeq 
where $p$ is an arbitrary momentum.
The first two terms in the sum are going to cancel the two propagators 
adjacent to the external gluon leg with momentum $q_i$, while  
the last one will contribute to the rest of the tensor structure in the 
numerator, now reduced by one power of $k$. 
 
\item When the dot product $(k \cdot \ep_i)$ appears in the numerator, 
we choose the four external gluon momenta
as our basis of Minkowski space. Hence we can expand the  external
polarization vector $\ep_i$ as
\beq 
\label{eq:ep_decomp} 
\ep_i^\mu = \sum_{j=1}^4 \ep_{ij} \, q_j^{\mu}, \qquad i=1\ldots 4, 
\eeq 
where the coefficients $\ep_{ij}$ are computed by inverting the
system of equations  
\beq 
\ep_i \cdot q_k = -\sum_{j=1}^4 \ep_{ij} \,\({\cal Q}_4\)_{jk}  , 
\qquad i,k=1\ldots 4, 
\eeq 
where the elements of the matrix ${\cal Q}_4$ are given by 
\beq 
\({\cal Q}_4\)_{jk} = - q_j \cdot q_k. 
\eeq 
Using Eq.~(\ref{eq:ep_decomp}), we can rewrite every scalar product of the 
form $(k\cdot \ep_i)$ in the numerator of the Feynman diagrams as a sum over  
$(k\cdot q_j)$, that we handle in the same way as shown in 
Eq.~(\ref{eq:k_dot_qi}).  

The matrix ${\cal Q}_4$  is singular when 
the four gluons become planar, i.e.\ when the four gluon momenta 
cease to be linearly independent.
This can easily be seen 
in the center-of-mass frame of partons 1 and 2, where we can write 
\beqn 
q_1 &=& E\(1,0,0,1\),\nonumber\\ 
q_2 &=& E\(1,0,0,-1\),\nonumber\\ 
q_3 &=& E_3\(1,\sin\th,0,\cos\th\),\nonumber\\ 
q_4 &=& E_4\(1,\sin\th'\cos\phi,\sin\th'\sin\phi,\cos\th'\),\nonumber 
\eeqn 
and the determinant of the matrix ${\cal Q}_4$ becomes 
\beq 
 \det {\cal Q}_4 = - 4 E^4 E_3^2 E_4^2 \(1-\cos^2\th'\) 
\(1-\cos^2\th\) \(1-\cos^2\phi\).  
\eeq 
The cuts imposed on the final-state partons (see Eq.~(\ref{eq:cuts_min}))  
avoid  
the singular region when one of the final-state partons is collinear 
with the initial-beam direction (singularities in $\th$ and $\th'$). 
 
The singularity in $\phi$ is un-physical, and the final amplitude should be 
finite  
near the singular points $\phi_S=0,\pi$. 
In our FORTRAN program, when the Monte Carlo integration approaches the 
singular points in $\phi$, we    
interpolate the value we need from the values of the amplitude in the points  
$\phi=\phi_S \pm 0.01 \,\pi$. 
We have checked that the interpolated amplitude differs from the exact value 
of the amplitude by less than $1\%$, in the non-singular region. 
\end{enumerate} 
 
By iterating this reduction procedure, we can write the contracted
amplitude for $gg\,\to\,ggH$ scattering in terms of
\begin{itemize} 
\item[-] twelve $E_{0}(p_a,p_b,p_c,p_d)$ functions, i.e.\ the scalar 
five-point functions computed in different kinematics; 
 
\item[-] twelve permutations of $D_{ij}$ functions with argument
$D_{ij}(p_a,p_b,p_c)$,   
with $a < c$, six $D_{ij}(p_a,p_b+p_c,p_d)$ with $a<d$ and twelve 
$D_{ij}(p_a,p_b,p_c+p_d)$, together with the corresponding $D_{0}$ functions; 
 
\item[-] three  $C_{ij}(p_a+p_b,p_c+p_d)$, and four 
$C_{ij}(p_a,p_b+p_c+p_d)$, together with the corresponding $C_{0}$ functions. 
 \end{itemize} 
As usual, the set $\{p_a,p_b,p_c,p_d\}$ is chosen in the group of permutations 
of the gluon momenta $\{q_1,q_2,q_3,q_4\}$.  To reduce the number of the 
$D_{ij}$ and $C_{ij}$ functions to this independent set, we made use of some 
identities between these functions, that we collect in 
Appendix~\ref{app:identities}.

As stated in Eq.~(\ref{eq:uncontract_gg}), we can study the color factors of
the $gg\,\to\,ggH$ process, by dividing the full
amplitude into three different classes, according to the number of fermionic
propagators in the loop.
We first  discuss the color factors of the diagrams containing a pentagon
loop.

\begin{itemize}
\item[-]{\bf Diagrams with a pentagon loop} \\
The contribution from the sum of charge-conjugated pentagon diagrams is 
proportional to the sum of two color traces with four $t$ matrixes (see 
Eq.~(\ref{eq:final_col_pent})). 
From the invariance property of the trace under cyclic 
permutations, we have only $(4-1)!=6$ independent 
traces, from the permutation of the four gluon indices.  
These six color traces combine together as in Eq.~(\ref{eq:final_col_pent}) to 
give rise to three independent color structures \beqn 
\label{eq:def_ci} 
c_1 &=& \Tr{t^{a_1} t^{a_2} t^{a_3} t^{a_4}}  + \Tr{t^{a_1} t^{a_4} t^{a_3}
t^{a_2}} \nonumber \\  
c_2 &=& \Tr{t^{a_1} t^{a_3} t^{a_4} t^{a_2}}  + \Tr{t^{a_1} t^{a_2} t^{a_4}
t^{a_3}}  \\  
c_3 &=& \Tr{t^{a_1} t^{a_4} t^{a_2} t^{a_3}}  + \Tr{t^{a_1} t^{a_3} t^{a_2}
t^{a_4}}. \nonumber  
\eeqn 
The $c_i$ color coefficients are real. In fact, using the identity 
(the sum over the repeated index is understood)
\beq 
\Tr{t^{a_1} t^{a_2} t^{a_3} t^{a_4}} = \frac{1}{4N}
\delta^{a_1a_2}\delta^{a_3a_4} +   
  \frac{1}{8}\(d^{a_1a_2l} + i \,f^{a_1a_2l}\) \(d^{a_3a_4l} + i
\,f^{a_3a_4l}\),  
\eeq 
where $f$ is the (totally antisymmetric) SU$(N)$ structure constant and 
$d$ is the totally symmetric symbol, 
we can write, for example, 
\beq 
\label{eq:def_c1}
c_1 = \frac{1}{4} \lq \frac{2}{N} \delta^{a_1a_2}\delta^{a_3a_4} + 
d^{a_1a_2l} d^{a_3a_4l} - f^{a_1a_2l} f^{a_3a_4l}\rq, 
\eeq 
and similar ones for $c_2$ and $c_3$.  Since $f$ and 
$d$ are real constants, the $c_i$ are real too. 
 
A few useful identities can be derived if we take the differences of the 
$c_i$
\beqn 
\label{eq:diff_ci} 
c_1 - c_2 = -\frac{1}{2} f^{a_1a_2l} f^{a_3a_4l} \quad\Longrightarrow \quad 
f^{a_1a_2l} f^{a_3a_4l} = 2\,(c_2-c_1), \nonumber\\ 
c_3 - c_1 = -\frac{1}{2} f^{a_1a_4l} f^{a_2a_3l} \quad\Longrightarrow \quad 
f^{a_1a_4l} f^{a_2a_3l} = 2\,(c_1-c_3),  \\ 
c_2 - c_3 = -\frac{1}{2} f^{a_1a_3l} f^{a_4a_2l} \quad\Longrightarrow \quad 
f^{a_1a_3l} f^{a_4a_2l} = 2\,(c_3-c_2). \nonumber 
\eeqn 
Note that the differences of the $c_i$ in the system~(\ref{eq:diff_ci})
automatically embodies the Jacobi identity:
by summing the three expressions, we have  
\beq 
f^{a_1a_2l} f^{a_3a_4l} + f^{a_1a_4l} f^{a_2a_3l} + f^{a_1a_3l} f^{a_4a_2l}
=0.  
\eeq

\item[-]{\bf Diagrams with a box loop}\\
These diagrams all contain a three-gluon vertex together with the quark
loop.
Since the sum of the charge-conjugated boxes is proportional to the structure
constant $f$ (see Eq.~(\ref{eq:final_col_box})), the final color factors
accompanying these diagrams are a product of two $f$'s, such as
$f^{a_1 a_2 l} f^{a_3 a_4 l}$.

With the help of Eq.~(\ref{eq:diff_ci}), we can express these products
in terms of differences of the $c_i$ color factors.
 
\item[-]{\bf Diagrams with a triangle loop} \\
The same argument can be used to show that the color structure of all the 
diagrams with a three-point function insertion are proportional to the 
product of two structure constants, that are then converted to
differences of $c_i$ color factors, using the identities of 
Eq.~(\ref{eq:diff_ci}).  
\end{itemize}

Since all the color structures of the diagrams contributing to the $gg\to 
ggH$ process can be written in terms of the $c_i$ color structures of 
Eq.~(\ref{eq:def_ci}), we can then decompose the full amplitudes in the 
following way  
\beq 
\label{eq:Agg_col_ci}
{\cal A}^{gg} = \sum_{i=1}^3 c_i \, {\cal A}^{gg}_i. 
\eeq  
The sum over the external colored gluons of the squared amplitude  
becomes 
\beq 
\sum_{\rm col} |{\cal A}^{gg}|^2 = \sum_{i,j=1}^3  {\cal A}^{gg}_i 
\({\cal A}^{gg}_j\)^* \sum_{\rm col} c_i \, c_j, 
\eeq 
where we have taken into account the fact that the $c_i$ are real (see
Eq.~(\ref{eq:def_c1})).
Using Eq.~(\ref{eq:def_ci}), one finds
\beqn 
{\cal C}_1 &\equiv& \sum_{\rm col} c_i \, c_i =  
\frac{\(N^2-1\) \(N^4-2N^2+6\)}{8N^2}, \qquad {\rm no\ summation\ over\ }i 
\\  
{\cal C}_2 &\equiv& \sum_{\rm col} c_i \, c_j =  
-\frac{\(N^2-1\) \(N^2-3\)}{4N^2}, \qquad i \neq j,  
\eeqn 
and we finally get 
\beq 
\sum_{\rm col} |{\cal A}^{gg}|^2 =  
{\cal C}_1  \sum_{i=1}^3  |{\cal A}^{gg}_i|^2  
+ {\cal C}_2  
\sum_{\stackrel{i,j=1}{i\neq j}}^3 {\cal A}^{gg}_i \({\cal A}^{gg}_j\)^*  
\eeq

\section{Checks}  
\label{sec:checks}
We were able to perform two different kinds of checks on the analytic 
amplitudes we computed: a gauge-invariance and a large-$\m$ limit check. 
 
\subsection{ Gauge invariance} 
\label{sec:gauge_invariance} 
 
Gauge invariance demands that the amplitudes should 
be invariant under the replacement $\ep_i\,\to\, \ep_i + \kappa_i q_i$, for
arbitrary values of $\kappa_i$.
This implies that for the $qg \,\to\, qgH$ process we must have (see 
Eq.~(\ref{eq:amp_qg}))  
\beq 
\label{eq:g_inv_qg} 
\left. 
\begin{array}[c]{c} 
q_{1}^{\al}\, \ep_{2}^{\bt}  
   \({\cal A}^{qg}\)_{\al\bt} =0\\ 
\ep_1^{\al}\, q_{2}^{\bt}  
   \({\cal A}^{qg}\)_{\al\bt} =0 
\end{array}  
\right\} \qquad {\rm when\ \ }  q_i \cdot \ep_i = 0, \quad i=1\ldots 2, 
\eeq 
and for $gg \,\to \,ggH$ (see Eq.~(\ref{eq:amp_gg})) 
\beq 
\label{eq:g_inv_gg} 
\left. 
\begin{array}[c]{c} 
q_{1}^{\al}\, \ep_{2}^{\bt} \, \ep_{3}^{\ga} \, \ep_{4}^{\de} \,   
   \({\cal A}^{gg}\)_{\al\bt\ga\de} =0\\ 
\ep_{1}^{\al}\, q_{2}^{\bt} \, \ep_{3}^{\ga} \, \ep_{4}^{\de} \,   
   \({\cal A}^{gg}\)_{\al\bt\ga\de} =0\\ 
\ep_{1}^{\al}\, \ep_{2}^{\bt} \, q_{3}^{\ga} \, \ep_{4}^{\de} \,   
   \({\cal A}^{gg}\)_{\al\bt\ga\de} =0\\ 
\ep_{1}^{\al}\, \ep_{2}^{\bt} \, \ep_{3}^{\ga} \, q_{4}^{\de} \,   
   \({\cal A}^{gg}\)_{\al\bt\ga\de} =0\\ 
\end{array}  
\right\} \qquad {\rm when\ \ }  q_i \cdot \ep_i = 0, \quad i=1\ldots 4\;. 
\eeq 
We checked the gauge invariance in Eqs.~(\ref{eq:g_inv_qg}) 
and~(\ref{eq:g_inv_gg}) both analytically and numerically (in the final 
Fortran program). 

Using Eq.~(\ref{eq:amp_qg}), it is straightforward to check that the
system~(\ref{eq:g_inv_qg}) is satisfied.
To check gauge invariance for the  $gg \,\to\, gg H$ process, 
we wrote the contracted amplitude of 
Eq.~(\ref{eq:amp_gg}) in terms of scalar pentagon ($E_0$), box ($D_0$), 
triangle ($C_0$) and bubble ($B_0$)  
integrals, keeping the space-time dimension $D$ arbitrary. 
This means that we expressed the $C_{ij}$ and $D_{ij}$ functions in 
terms of $B_0$, $C_0$ and $D_0$ ``master'' integrals.  
The coefficients of these scalar integrals are then functions of scalar
products  
$(q_i \cdot q_j)$, $(q_i \cdot \ep_j)$,  $(\ep_i \cdot \ep_j)$ and of the 
$\ep_{ij}$ coefficients introduced in Eq.~(\ref{eq:ep_decomp}). 
 
Since the two-point functions $B_0$ are divergent in $\ep=(D-4)/2$, 
and since the total amplitude must be finite, these poles must cancel. 
In fact, the factors multiplying  the $B_0$ contributions are proportional to 
$(D-4)$, so that only the pole coefficient of $B_0$ contributes to the 
finite amplitude (see comment after Eq.~(\ref{eq:ft})).

To implement the gauge invariance check  with 
respect to the polarization vector $\ep_i$, as 
described in the system~(\ref{eq:g_inv_gg}),  we make the replacement 
\beq 
\ep_i \,\to\, q_i \quad \Longrightarrow \quad 
\ep_{ii} = 1, \qquad \ep_{ij} = 0, \quad j \neq i, 
\eeq 
in Eq.~(\ref{eq:amp_gg}), 
and we impose the orthogonality condition $\ep_k \cdot q_k = 0$, that 
constrains the $\ep_{kj}$ coefficients to satisfy the identity  
(see Eq.~(\ref{eq:ep_decomp})) 
\beq 
\ep_k \cdot q_k = \sum_{j=1}^4 \ep_{kj} \, \(q_j \cdot q_k\) = 0,  
\qquad k \neq i. 
\eeq

We have checked gauge invariance in two different ways. 
\begin{enumerate} 
\item Suppose that instead of considering the QCD process $gg\,\to\, ggH$, we 
 consider the QED analogue, $\ga\ga\,\to\, \ga\ga H$. 
 In this scenario, all the diagrams with a three- or a four-gluon vertex are 
 no longer present: 
 the only surviving diagrams are the ones containing a  
 pentagon loop, with no color structure associated. 
 The amplitude, not contracted with any external photon polarization vectors, 
 is (see Eq.~(\ref{eq:uncontract_gg}) for comparison) 
 \beq 
 \label{eq:uncontract_gamma} 
 \({\cal A}^{\ga \ga\,\to\, \ga \ga H}\)_{\al\bt\ga\de}  
   \equiv \sum_{i=1}^{12} P^i_{\al\bt\ga\de}. 
 \eeq 
 The gauge invariance of this expression allows us to check the correctness
 of the tensor reduction of the pentagon diagrams only.  We have contracted
 Eq.~(\ref{eq:uncontract_gamma}) with the polarization vectors of the photons
 and we have applied the tensor reduction procedure previously
 described. Instead of expressing the results in terms of $C_{ij}$ and
 $D_{ij}$ functions, we have expressed these coefficient functions in terms
 of $B_0$, $C_0$, $D_0$ and $E_0$ scalar integrals, keeping the space-time
 dimension $D$ arbitrary.  Since these scalar integrals form a set of
 independent functions, we expect the coefficients of these integrals to be
 zero, in order to fulfill the gauge-invariance test.  Note that we have
 considered the twelve scalar five-point functions $E_0$ as independent from
 the four-point functions $D_0$, that is we have not used
 Eq.~(\ref{eq:pent_fun_boxes}).  This is indeed the case, since, in arbitrary
 $D$ dimensions, the scalar pentagon cannot be expressed as a combination of
 scalar boxes only, so that it is really an independent integral.

\item Finally, we have checked that our full QCD amplitude satisfies the four
identities in the system~(\ref{eq:g_inv_gg}).  Since the amplitude can be
split into three different contributions according to the three independent
color factors $c_i$ (see Eq.~(\ref{eq:Agg_col_ci})), this means that not only
the full amplitude is gauge invariant, but that the
three sub-amplitudes ${\cal A}_i^{gg}$ are separately gauge invariant, and
satisfy a system of equations similar to~(\ref{eq:g_inv_gg}).
\end{enumerate}

\subsection{\boldmath Large-$\m$ limit}  
The amplitudes for Higgs plus two partons agree in the large-$\m$ limit with
the corresponding amplitudes obtained from the heavy-top effective
Lagrangian~\cite{kauffman}.  This check was done numerically by setting 
$\m=3$~TeV. We found good agreement with the $\m\,\to\,\infty$ results, 
within the statistical errors of the Monte Carlo program, for Higgs boson 
masses in the range 100~GeV $< m_H <$ 700~GeV.

\section{Applications to LHC physics}
\label{sec:pheno}
The gluon-fusion processes at ${\cal O}(\alpha_s^4)$, together with 
weak-boson fusion ($qq\,\to\, qqH$ production via $t$-channel exchange of a $W$
or $Z$), 
are expected to be the dominant sources of $H+2$~jet events at the LHC. The
impact of the former on LHC Higgs phenomenology is determined by the relative 
size of these two contributions. However, the gluon-fusion cross sections 
for $H+2$~jet events diverges as the final-state partons become collinear
with one another or with the incident beam directions, or as final-state
gluons become soft. A minimal set of cuts on the final-state partons, which
anticipates LHC detector capabilities and jet finding algorithms, is required
to define an $H+2$~jet cross section. Our minimal set of cuts is
\bq \label{eq:cuts_min}
p_{Tj}>20\;{\rm GeV}, \qquad |\eta_j|<5,\qquad R_{jj}>0.6,
\eq
where $p_{Tj}$ is the transverse momentum of a final state parton
and $R_{jj}$ describes the separation of the two partons in the 
pseudo-rapidity $\eta$ versus azimuthal angle plane
\beq 
R_{jj} = \sqrt{\Delta\eta_{jj}^2 + \phi_{jj}^2}\;.
\eeq

\begin{figure}[htb]
\centerline{\epsfig{figure=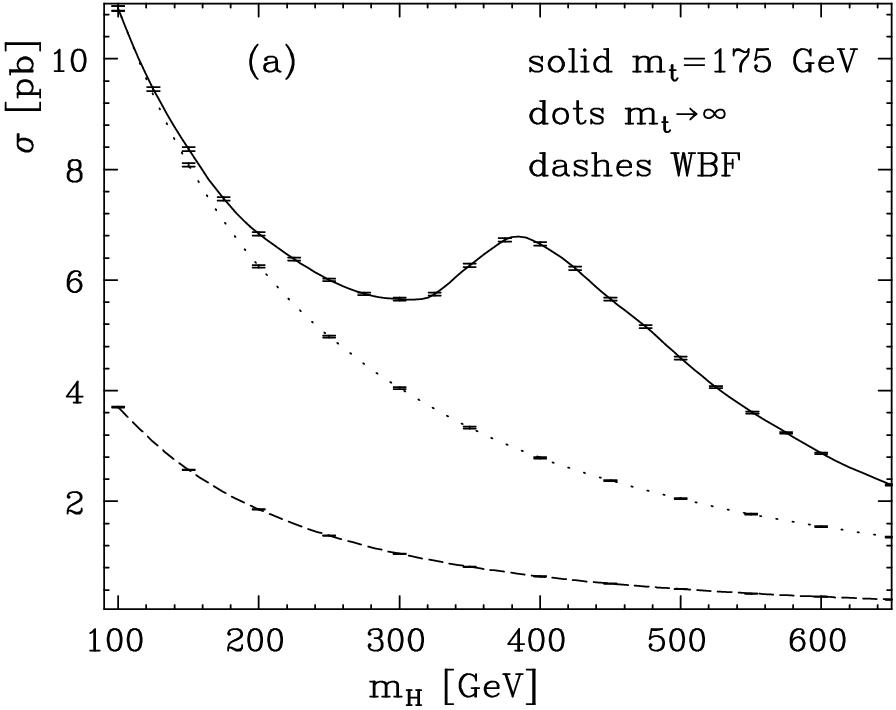,width=0.495\textwidth,clip=}
\epsfig{figure=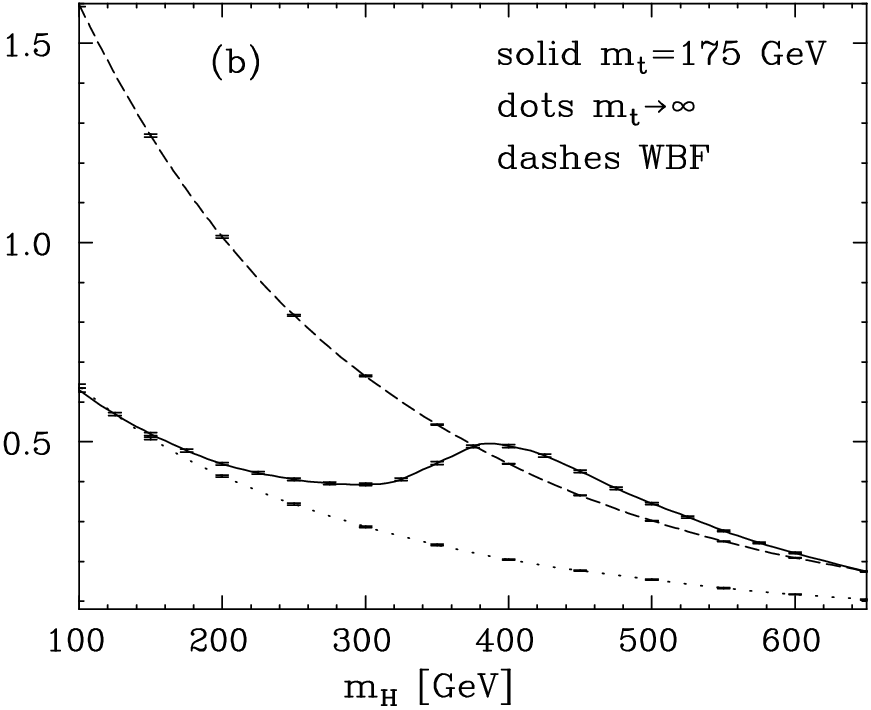,width=0.48\textwidth,clip=}
}
\ccaption{}
{ \label{fig:sigmaMh} 
$H+2$~jet cross sections in pp collisions at 
{\rm $\protect\sqrt{s}=14$~TeV} as a function of the Higgs boson mass.
Results are shown for gluon-fusion processes induced by a top-quark loop
with {\rm $\m=175$~GeV} and in the $\m\,\to\,\infty$ limit, computed using the
heavy-top effective Lagrangian, 
and for weak-boson fusion. The two panels correspond to two sets of jet cuts:
(a) inclusive selection (see 
Eq.~(\protect\ref{eq:cuts_min})) and (b) WBF selection 
(Eqs.~(\protect\ref{eq:cuts_min}) and~(\protect\ref{eq:cut_gap})).}
\end{figure}

\begin{figure}[htb]
\centerline{\epsfig{figure=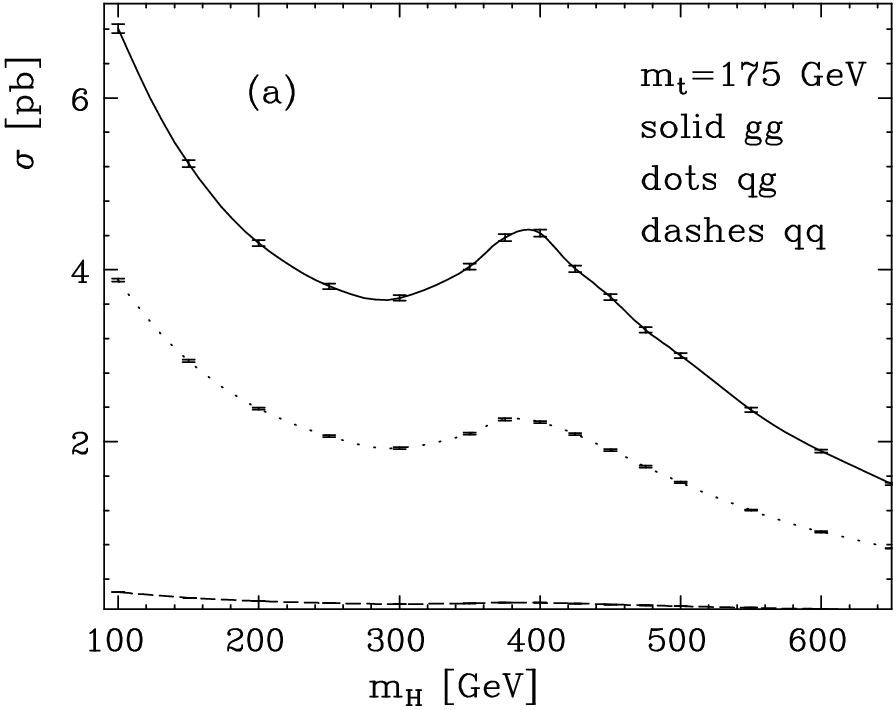,width=0.495\textwidth,clip=}
\epsfig{figure=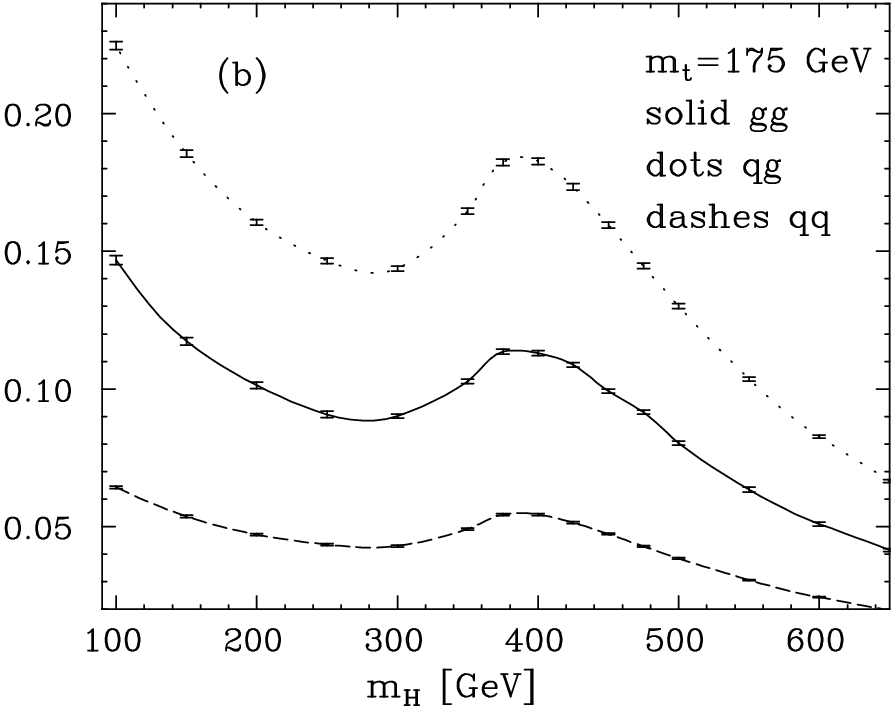,width=0.494\textwidth,clip=}
}
\ccaption{}
{ \label{fig:sigma_contribs} 
$H+2$~jet contributions to the cross section in pp collisions at 
{\rm $\protect\sqrt{s}=14$~TeV} as a function of the Higgs boson mass.
Results are shown for the different contributions to the gluon-fusion process 
($gg, qg$ and $qq$ amplitudes) using (a) the inclusive cuts of 
Eq.~(\protect\ref{eq:cuts_min}) and (b) the WBF cuts of 
Eqs.~(\protect\ref{eq:cuts_min}) and~(\protect\ref{eq:cut_gap}).
} 
\end{figure}

Expected $H+2$~jet cross sections at the LHC are shown in
Fig.~\ref{fig:sigmaMh}, as a function of the Higgs boson mass, $m_H$.  The
three curves compare results for the expected SM gluon-fusion cross section
at $\m=175$~GeV (solid line) with the large-$\m$ limit (dotted line), and
with the WBF cross section (dashed line).
Error bars indicate the statistical errors of the Monte Carlo integration. 
Cross sections correspond to the sum over all Higgs decay modes: finite Higgs
width effects are included.

In all our simulations, we used the CTEQ4L set for parton-distribution
functions~\cite{cteq4}. Unless specified otherwise, the factorization scale
was set to $\mu_f=\sqrt{p_{T1} \, p_{T2} }$. Since this calculation 
is a LO one,
we employ one-loop running of the strong coupling constant. In
Fig.~\ref{fig:sigmaMh} we fix $\alpha_s = \alpha_s(M_Z) =0.12$.

The left panel in Fig.~\ref{fig:sigmaMh} shows cross sections within the
minimal cuts of Eq.~(\ref{eq:cuts_min}). The gluon-fusion contribution 
dominates because the cuts retain events with jets in the central region, 
with relatively small dijet invariant mass. In order to assess background 
levels for WBF events, it is more appropriate to consider typical tagging 
jet selections employed for WBF studies~\cite{RZ_WW}. This is done in 
Fig.~\ref{fig:sigmaMh} (b) where, in addition to the cuts of 
Eq.~(\ref{eq:cuts_min}), we require
\bq \label{eq:cut_gap}
|\eta_{j1}-\eta_{j2}|>4.2, \qquad \eta_{j1}\cdot\eta_{j2}<0, \qquad
m_{jj}>600\;{\rm GeV},
\eq
i.e.\ the two tagging jets must be well separated, 
they must reside in 
opposite detector hemispheres and they must possess a large dijet 
invariant mass. With these selection cuts the weak-boson 
fusion processes dominate over gluon fusion by about 3/1 for 
Higgs boson masses in the 100 to 200~GeV range. This means that a relatively
clean separation of weak-boson fusion and gluon-fusion processes will be 
possible at the LHC, in particular when extra central-jet-veto techniques are 
employed to further suppress semi-soft gluon radiation in QCD backgrounds.
We expect that a central-jet veto will further suppress gluon fusion with
respect to WBF by an additional factor of three~\cite{RZ_WW}.

A conspicuous feature of the $H+2$~jet gluon-fusion cross sections in 
Fig.~\ref{fig:sigmaMh} is the threshold enhancement at $m_H\approx 2\,\m$,
an effect which is familiar from the inclusive gluon-fusion cross section.
Near this ``threshold peak'' the gluon-fusion cross section rises to equal
the WBF cross section, even with the selection cuts of Eq.~(\ref{eq:cut_gap}).
Well below this region, the large $\m$ limit provides an excellent 
approximation to the total $H+2$~jet rate from gluon fusion, at least when 
considering the total Higgs production rate only. Near top-pair threshold
the large $\m$ limit underestimates the rate by about a factor of 2.

\begin{figure}[hbt]
\centerline{\epsfig{figure=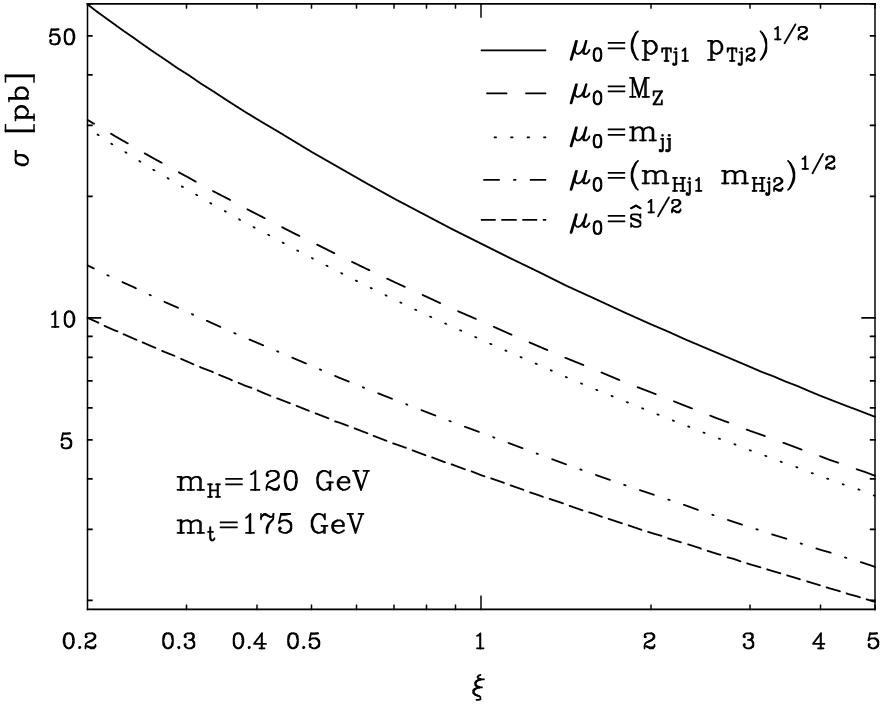,width=0.75\textwidth,clip=}}
\ccaption{} { \label{fig:ren_scale} Renormalization-scale dependence of the
total cross section for $H$ plus two jet production with the inclusive cuts
of Eq.~(\protect\ref{eq:cuts_min}). The renormalization 
scale $\mu_r=\xi \mu_0$  is varied in the range $1/5 < \xi < 5$.
The five curves correspond, from top to bottom, to the following choice of
$\mu_0$: the geometric mean
of the transverse momenta of the two jets, the $Z$ mass, the
invariant mass of the two jets, the geometric mean of the two invariant
masses of the Higgs and the jets, and the partonic center-of-mass energy.
}
\end{figure}

A somewhat surprising feature of Fig.~\ref{fig:sigmaMh} (b) is the excellent
approximation provided by the large $\m$ limit at Higgs boson masses below
about 200~GeV. Naively one might expect the large dijet invariant mass,
$m_{jj}>600$~GeV, and the concomitant large parton center-of-mass energy 
to spoil the $\m\,\to\,\infty$ approximation. This is not the case, however.
As shown in Ref.~\cite{DKOSZ}, the large $\m$ limit works well in the 
intermediate Higgs mass range, as long as jet 
transverse momenta stay small: $p_{Tj}\lsim \m$.

In Fig.~\ref{fig:sigma_contribs} we have plotted the
individual contributions to the gluon-fusion cross section which  are
coming from the $gg\,\to\, ggH$, $qg\,\to\, qgH$ and 
$qq\,\to\, qqH$ sub-processes, including all crossed processes for each of the 
three subgroups. Results are shown after imposing the inclusive cuts of 
Eq.~(\protect\ref{eq:cuts_min})  (left panel) and the WBF cuts of 
Eqs.~(\protect\ref{eq:cuts_min}) and~(\protect\ref{eq:cut_gap}) (right
panel), with $\m=175$~GeV, so that the sum of the three curves in each panel
add up to the solid line curve in Fig.~\ref{fig:sigmaMh}.
External gluons dominate in the inclusive-cut case (left panel):
final-state gluons
tend to be soft and initial gluons preferably lead to soft events due to the 
rapid fall-off of the gluon parton distribution function, $g(x,\mu_f)$,
with increasing $x$.
When the $m_{jj}>600$~GeV constraint of the WBF cuts is imposed, the gluon
contribution dies rapidly, as is shown in Fig~\ref{fig:sigma_contribs} (b).

\begin{figure}[t]
\centerline{\epsfig{figure=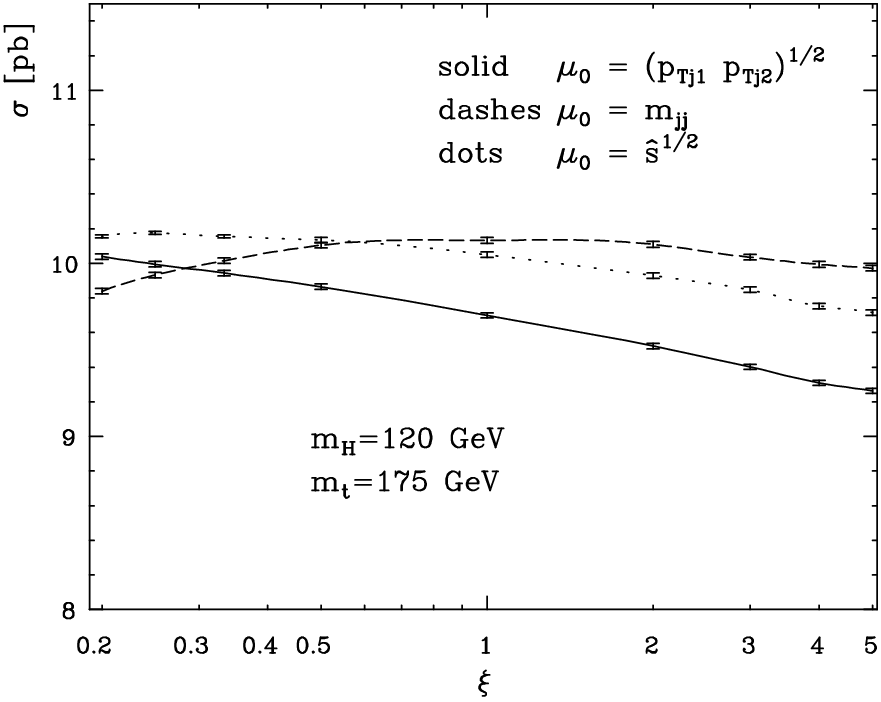,width=0.75\textwidth,clip=}}
\ccaption{} { \label{fig:fac_scale} Factorization-scale dependence of the
total cross section for $H$ plus two jet production with the inclusive cuts
of Eq.~(\protect\ref{eq:cuts_min}). The
factorization
scale $\mu_f=\xi \mu_0$  is varied in the range $1/5 < \xi < 5$.
The three curves correspond to the following choice of
$\mu_0$: the geometric
average between the transverse momenta of the two jets, the 
invariant mass of the two jets and the partonic center-of-mass energy.}
\end{figure}

The results shown in Fig.~\ref{fig:sigmaMh} raise two questions, which we
intend to answer in the following: i) what are the uncertainties in the 
prediction of the $H+2$~jet cross section, and ii) which distributions 
are the most effective in distinguishing gluon-fusion and weak-boson fusion
contributions to $H+2$~jet events?\\
In order to asses the sensitivity of the gluon-fusion cross section to 
higher order QCD corrections, we have plotted 
in Fig.~\ref{fig:ren_scale}  the total cross section for
several choices of the renormalization scale (the factorization scale has
been kept at $\mu_f=\sqrt{p_{T1} \, p_{T2} }$).
We have fixed $\Lambda_{\rm
\overline{MS}}^5= 254$~MeV, so that, $\as(M_Z)=0.12$, with $n_f=5$ active
flavors. 
We have chosen five different scales: $\mu_0=\sqrt{p_{T1}\, p_{T2}},\ M_Z,\
m_{jj},\ \sqrt{m_{Hj1} \,m_{Hj2}}$ and $\sqrt{\hat{s}}$, i.e.\ the geometric
average between the transverse momenta of the two jets, the $Z$ mass, the
invariant mass of the two jets, the geometric mean of the two invariant
masses of the Higgs and the jets and the partonic center-of-mass energy.
For every event generated by our Monte Carlo, we have computed the running of
the coupling constant 
$\as(\mu_r)$ at the values $\mu_r = \xi \mu_0$, where $\xi$ was allowed to vary
from 1/5 to 5.
We can see that the renormalization-scale dependence is very strong, mainly
due to the fact that this is a leading-order calculation, at order $\as^4$.

What is the ``most natural'' scale for $\as$ is an unresolved issue.  The
good agreement~\cite{DKOSZ}
 between the complete result and the $\m\,\to\,\infty$ one
(away from threshold) implies that the cross section is dominated by Feynman
diagrams with a gluon exchange in the $t$ channel. These diagrams contain a
triangle loop that couples the $t$-channel gluon with the Higgs.
For this reason, it seems reasonable to make the replacement
\beq
\label{eq:as_true}
\as^4 \, \to \, \as(p_{T1})\,\as(p_{T2}) \,\as^2(m_H).
\eeq
With this choice for the strong coupling constant we have a total cross 
section (within the cuts of Eq.~(\ref{eq:cuts_min})) of about $\sigma=9.6$~pb, 
which sits in between the two values computed using
$\as^4(M_Z)$ and $\as^4(m_{jj})$ (see Fig.~\ref{fig:ren_scale}).
Dismissing the extreme choice $\mu_r=\sqrt{\hat s}$ (which is ill defined 
at higher orders), and allowing for the conventional factor of 2 variation
of $\xi$, Fig.~\ref{fig:ren_scale} suggests an uncertainty of the gluon fusion
$H+2$~jet cross section of about a factor 2.5 as compared to the central value
of 9.6~pb obtained with the central choice of Eq.~(\ref{eq:as_true}).

Keeping fixed the renormalization scale at $M_Z$, we have collected 
in Fig.~\ref{fig:fac_scale} the results for the
factorization-scale dependence of the total cross section.
The factorization scale $\mu_f=\xi \mu_0$ was  allowed to vary in the range
described by $1/5 < \xi < 5$, where $\mu_0$ was taken equal to the geometric
average of the transverse momenta of the two jets, the
invariant mass of the two jets and the partonic center-of-mass energy.
Compared with the variation with respect to the renormalization scale, we
see that the dependence on the factorization scale is almost negligible:
in Fig.~\ref{fig:fac_scale} the $H+2$~jet cross section varies
between 9.2~pb and 10.2~pb.

In the following we use the renormalization scale choice of 
Eq.~(\ref{eq:as_true}) and set $\mu_f=\sqrt{p_{T1} \, p_{T2} }$. We take 
$m_H=120$~GeV throughout, as a characteristic Higgs boson mass. 

\begin{figure}[htb]
\centerline{ 
\epsfig{figure=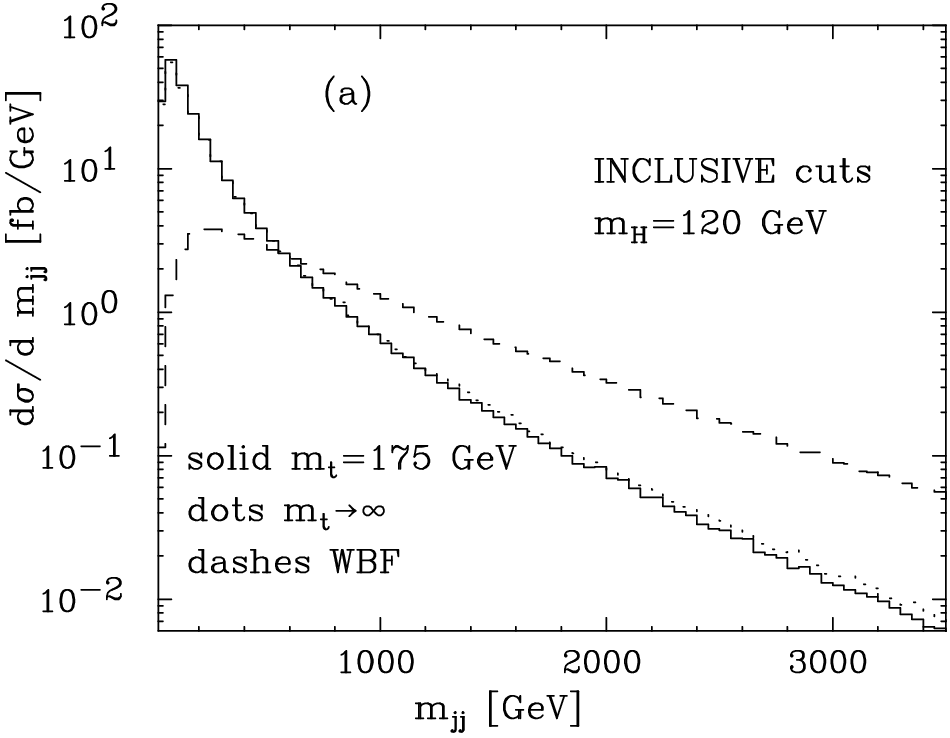,width=0.515\textwidth,clip=}\ \
\epsfig{figure=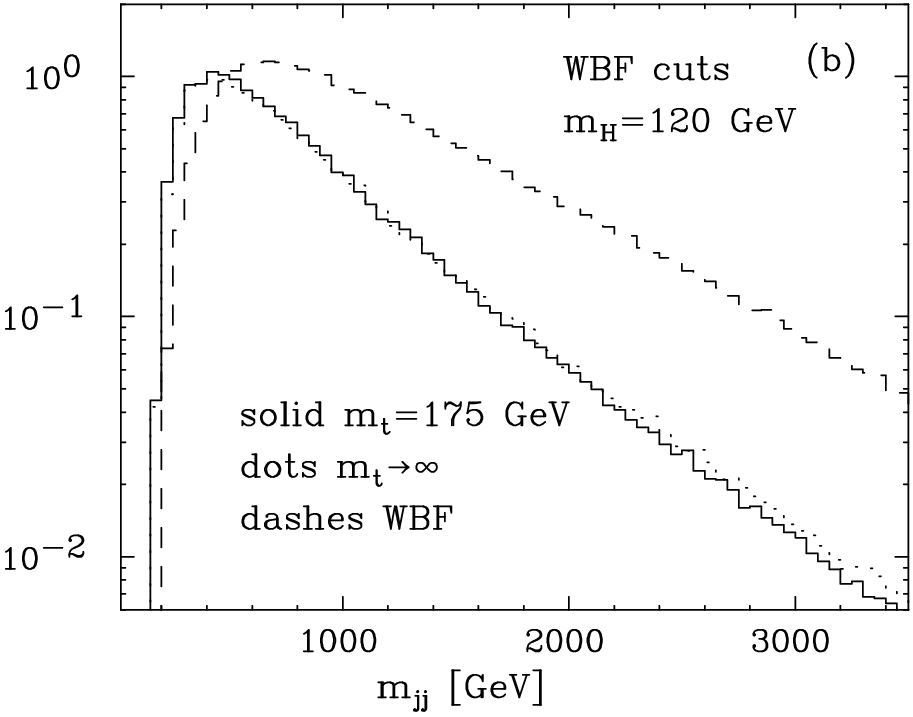,width=0.495\textwidth,clip=}}
\ccaption{} { \label{fig:mjj_comb} Dijet invariant-mass distribution of the
two final jets for gluon-fusion (solid) and WBF (dashes) processes. 
Left panel (a): inclusive cuts of Eq.~(\protect\ref{eq:cuts_min}); right
panel (b) WBF cuts of Eqs.~(\protect\ref{eq:cuts_min})
and~(\protect\ref{eq:cut_gap}), where we have suppressed the constraint
{\rm $m_{jj}>600$~GeV}.}
\end{figure}

\begin{figure}[htb]
\centerline{
\epsfig{figure=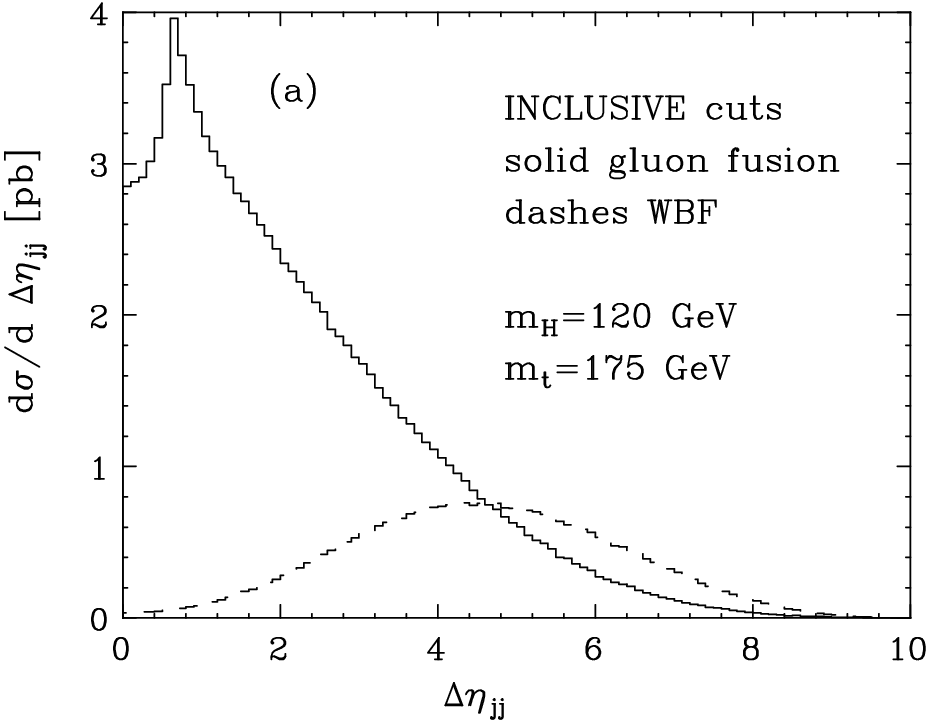,width=0.52\textwidth,clip=}\ \ 
\epsfig{figure=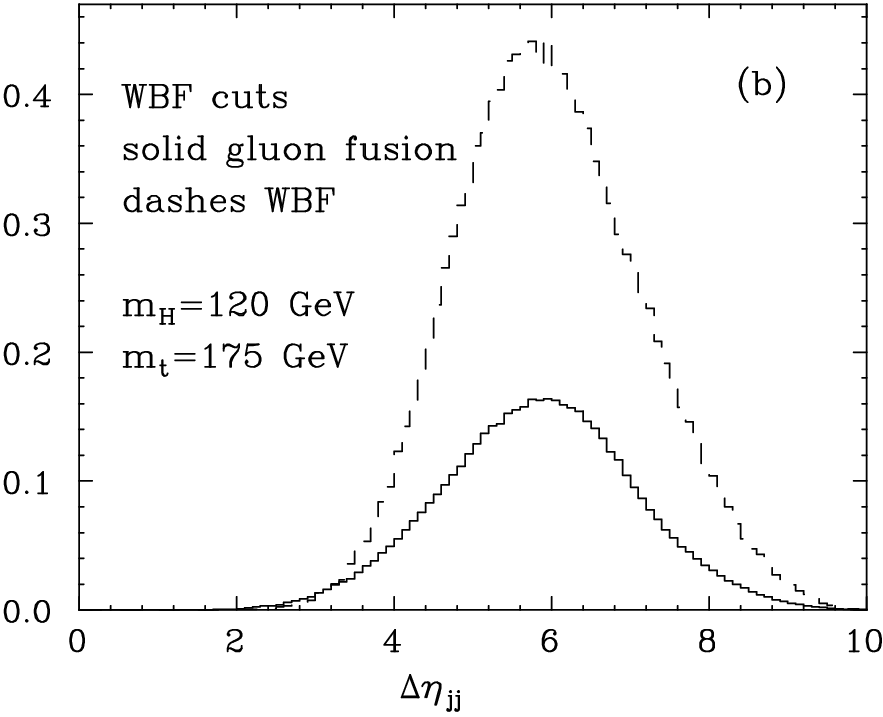,width=0.495\textwidth,clip=}}
\ccaption{} { \label{fig:eta_jj_inclusive} Rapidity separation of the
two final jets for gluon-fusion (solid) and WBF (dashes) processes. 
Left panel (a): inclusive cuts of Eq.~(\protect\ref{eq:cuts_min}); right
panel (b) WBF cuts of Eqs.~(\protect\ref{eq:cuts_min})
and~(\protect\ref{eq:cut_gap}), where we have suppressed the constraint
$|\eta_{j1}-\eta_{j2}|>4.2$.}
\end{figure}

Turning now to the issue of differentiating between
gluon fusion and WBF processes, the prominent characteristics to be considered 
here are the jet properties.
Figure~\ref{fig:mjj_comb} shows the dijet-mass
distribution for gluon-fusion and WBF processes, using the inclusive 
and the WBF cuts. In this last case, we have suppressed the constraint
{\rm $m_{jj}>600$~GeV}, in order to access the region of small dijet
invariant mass.
In both panels, the high dijet mass region ($m_{jj}> 1$~TeV) is dominated by
WBF. The significantly softer dijet-mass spectrum of the gluon-fusion 
processes is characteristic of QCD processes, which are dominated by
external gluons, as compared to quarks in WBF processes (see
Fig.~\ref{fig:sigma_contribs} and comments about it).
Figure~\ref{fig:mjj_comb} also shows the $\m\,\to\,\infty$ dijet-mass
distributions (dotted curves), that are almost indistinguishable from the 
$\m=175$~GeV result:
large dijet invariant masses do not invalidate the
$\m\,\to\,\infty$ limit as long as the Higgs boson mass and the jet transverse
momenta are small enough, less than $\m$ in practice.

A characteristic of WBF events is the large rapidity separation 
of the two tagging jets, a feature which is not shared by $H+2$~jet events 
arising from gluon fusion. The rapidity separation of the jets is shown 
in Fig.~\ref{fig:eta_jj_inclusive}, for both gluon-fusion (solid) and
WBF (dashes) processes. The two panels correspond to the inclusive cuts 
of Eq.~(\ref{eq:cuts_min}) and to the stricter WBF cuts of 
Eq.~(\ref{eq:cut_gap}), where we have suppressed the constraint 
$|\eta_{j1}-\eta_{j2}|>4.2$, in order to have access to the entire 
$\Delta\eta_{jj}$ range. 
The jet separation cut, $|\eta_{j1}-\eta_{j2}|>4.2$,
is one of the most effective means of enhancing WBF processes with respect to
gluon fusion.
The small dip in the gluon-fusion distribution at small $\Delta\eta_{jj}$
is a consequence of the cut $R_{jj}> 0.6$.

\begin{figure}[htb]
\centerline{\epsfig{figure=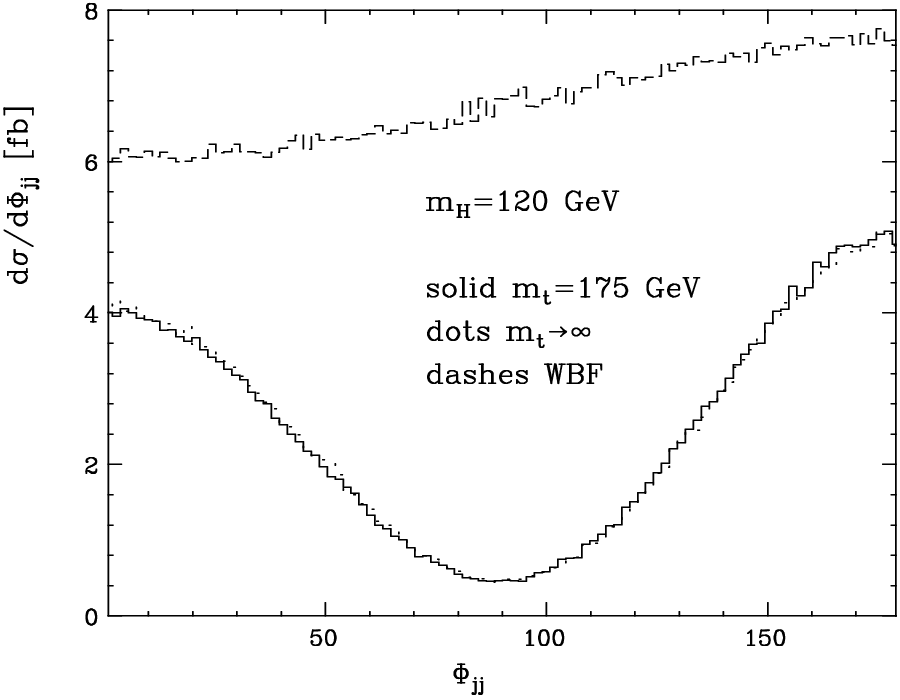,width=0.75\textwidth,clip=}}
\ccaption{} { \label{fig:phi_cuts_comp} Azimuthal-angle distribution between
the two final jets, with the WBF cuts of Eqs.~(\protect\ref{eq:cuts_min})
and~(\protect\ref{eq:cut_gap}). 
Results are shown for gluon-fusion processes induced by a top-quark loop
with $\m=175$~{\rm GeV} and in the $\m\,\to\,\infty$ limit, computed using the
heavy-top effective Lagrangian, and for weak-boson fusion.}
\end{figure}

A second jet-angular correlation, which allows to distinguish
gluon fusion from weak-boson fusion, is the azimuthal angle between the two
jets, $\phi_{jj}$. The distributions for gluon-fusion and WBF processes are
shown in Fig.~\ref{fig:phi_cuts_comp}. 
In the WBF process $qQ \,\to\, qQH$, the matrix element squared is
proportional to
\beq
  |{\cal A}_{\rm WBF}|^2 \propto
\frac{1}{\(2 \,\overline p_1\cdot \overline p_2+M_W^2\)^2}\,
\frac{1}{\(2 \,\overline p_3\cdot \overline p_4+M_W^2\)^2}
\, \hat{s} \, m_{jj}^2\;,
\eeq
and is dominated by the contribution in the forward region, where the
dot products in the denominator are small.
Since the dependence of $m_{jj}^2$ on $\phi_{jj}$ is mild, we have the flat
behavior depicted in Fig.~\ref{fig:phi_cuts_comp}.
The azimuthal-angle distribution of the gluon-fusion process is instead
characteristic of the CP-even operator $H G_{\mu\nu}G^{\mu\nu}$, where
$G_{\mu\nu}$ is the gluon field strength tensor~\cite{PRZ}. 
This effective coupling
can be taken as a good approximation for the $ggH$ coupling in the 
high-$\m$ limit. Note that the large-$\m$ limit (dotted line) is almost
indistinguishable from the $\m=175$~GeV result (solid line).

\begin{figure}[htb]
\centerline{\epsfig{figure=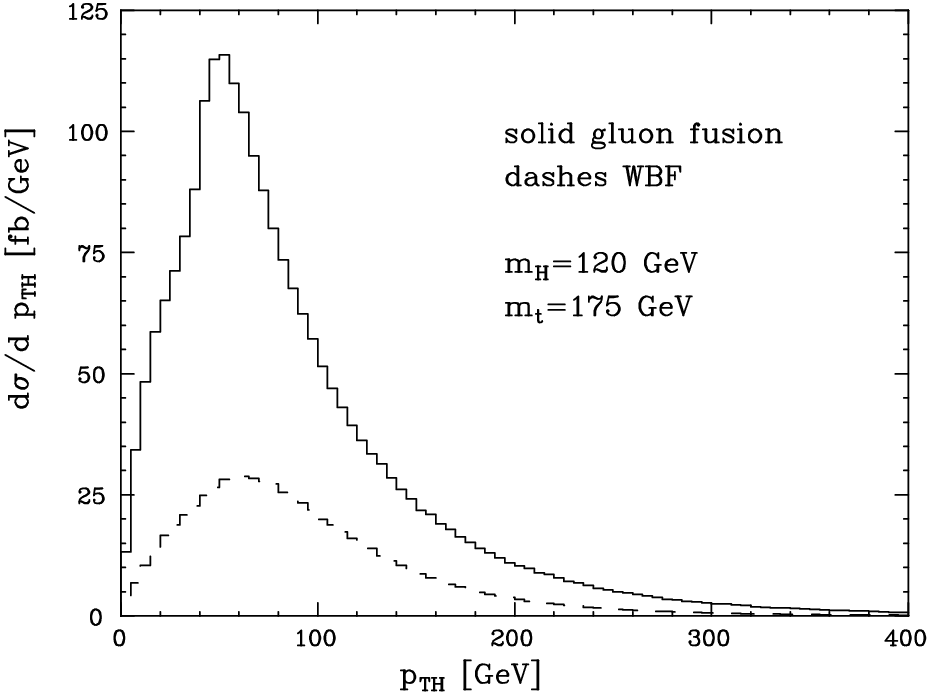,width=0.75\textwidth,clip=}}
\ccaption{} { \label{fig:pt_H_inclusive}  Transverse-momentum distribution of
the Higgs from gluon-fusion (solid) and from WBF (dashes) processes with the
inclusive selection of Eq.~(\protect\ref{eq:cuts_min}). }
\end{figure}

Finally, in Fig.~\ref{fig:pt_H_inclusive}, we show the transverse-momentum
distribution of the Higgs boson in gluon-fusion (solid lines) and in WBF
(dashes lines) processes, with the inclusive selection of 
Eq.~(\protect\ref{eq:cuts_min}).  Within these cuts, both 
differential cross sections peak around a value of $p_{TH} \approx 50$~GeV.
Note, however, that, while the peak position of the WBF distribution 
is largely tied to the mass of the exchanged intermediate weak bosons, the 
peaking of the gluon fusion processes occurs just above 40~GeV, which is a 
direct consequence of the $p_{Tj}>20$~GeV cut of Eq.~(\ref{eq:cuts_min}).

\section{Conclusions} 
\label{sec:concl}

In the previous sections, we have provided the results of the ${\cal
O}(\alpha_s^4)$ calculation of $H+2$~jet cross section, including the full
top-mass dependence. For the quark-quark and quark-gluon scattering
amplitudes we have found very compact analytic expressions. The $gg\to ggH$
amplitudes, which include pentagon loops, are more complex and available
analytically, as {\tt Maple} output, and numerically, in the form of a
FORTRAN program.

Numerical investigations of the resulting cross sections at the LHC 
provide  many interesting insights. With minimal jet-selection cuts (parton
separation of $R_{jj}>0.6$ and jet transverse momenta in excess of 20~GeV)
the gluon-fusion induced cross section is sizable, of order 10~pb for
$m_H=120$~GeV, which corresponds to about 30\% 
of the inclusive Higgs production rate.
Since our calculation gives the $H+2$~jet rate at leading order, it 
exhibits the large renormalization-scale dependence to be expected of an 
order $\alpha_s^4$ process.  

As expected, the large-$\m$ limit provides an excellent approximation to the
full $\m$ dependence when the Higgs mass is small compared to the top-pair 
threshold.
The large-$\m$ limit is found to break down for $m_H>\m$ 
and when jet transverse momenta become large ($p_{Tj}\gsim\m$). However,
large dijet invariant masses do not invalidate the 
$\m\,\to\,\infty$ 
limit, as long as the Higgs boson mass and the jet transverse momenta are
small enough, less than the top-quark mass in practice. 
This observation opens the possibility of NLO corrections to $H+2$~jet 
production from gluon fusion. Performing the calculation in the large-$\m$
limit would correspond to a 1-loop calculation of a $2\,\to\, 3$ process. 
Such a calculation might be desirable to reduce systematics errors in the 
extraction of $HZZ$ and $HWW$ couplings from the competing weak-boson 
fusion processes at the LHC.

Consideration of the gluon fusion $H+2$~jet rate as a background to WBF studies
constitutes an important application of our calculation. While the overall
$H+2$~jet rate is dominated by gluon fusion at the LHC, kinematic properties 
of the two processes are sufficiently different to allow an efficient 
separation. Gluon-fusion events tend to be soft, with a relatively small 
separation of the two jets. In contrast, the two tagging jets of weak-boson
fusion events have very large dijet invariant mass, and are far separated
in rapidity. Using rapidity and invariant-mass cuts, the gluon-fusion 
cross section can be suppressed well below the WBF rate. In addition,
the azimuthal angle between the two jets shows a dip at 90 degrees which
is characteristic for loop-induced Higgs couplings to gauge 
bosons~\cite{PRZ}.  Based on our calculation we conclude that a 
relatively clean separation of weak-boson fusion and gluon fusion Higgs 
plus two-jet events will be possible at the LHC.

\subsection*{Acknowledgments} 
We thank E.~Richter-Was  for insisting on the importance
of this calculation at an early stage.
C.S. acknowledges the U.S. National Science Foundation under grant 
PHY-0070443. 
W.K. acknowledges the DOE funding under Contract  
No.~DE-AC02-98CH10886.  
This research was supported in part by the University of Wisconsin Research 
Committee with funds granted by the Wisconsin Alumni Research Foundation and 
in part by the U.~S.~Department of Energy under Contract 
No.~DE-FG02-95ER40896.

\appendix
\section{Scalar integrals: \boldmath $C_0, D_0$ and 
	$E_0$ functions} 
\label{app:CDE}
All the scalar integrals needed for the calculation  
are finite in $D=4$ dimensions, due to the presence of the top-quark mass. 
No further regulator is required. 
Scalar triangles ($C_0$) and boxes ($D_0$) have been known for a long time in
the  
literature~\cite{tris_boxes} and efficient computational procedures are 
available~\cite{Denner}. 
Following the procedure outlined in Refs.~\cite{BDK}, we can 
express all scalar pentagons as linear combinations of scalar boxes 
\beq 
\label{eq:pent_fun_boxes} 
 E_0(p_1,p_2,p_3,p_4) = \sum_{i,j=1}^{5} {\cal F}_{ij}\,  D_0^j, 
\eeq 
where 
\beqn 
D_0^1 &=& D_0\(p_2,p_3,p_4\) \nonumber \\ 
D_0^2 &=& D_0\(p_4,p_3,p_1+p_2\) \nonumber \\ 
D_0^3 &=& D_0\(p_1,p_2+p_3,p_4\)  \\ 
D_0^4 &=& D_0\(p_1,p_2,p_3+p_4\) \nonumber \\ 
D_0^5 &=& D_0\(p_1,p_2,p_3\), \nonumber  
\eeqn 
and  the matrix ${\cal F}={\cal C}^{-1}$, with 
\beq 
{\cal C}_{ij} = (r_i-r_j)^2 - 2\, \m^2, 
\eeq 
and 
\beqn 
 r_1 &=& 0 \nonumber \\ 
 r_2 &=& p_1 \nonumber \\ 
 r_3 &=& p_{12} = p_1+p_2\\ 
 r_4 &=& p_{123}= p_1+p_2+p_3 \nonumber \\ 
 r_5 &=& p_{1234}=p_1+p_2+p_3+p_4 \;. \nonumber  
\eeqn 
 
\section{\boldmath Relations among $C_{ij}$ and $D_{ij}$ functions} 
\label{app:identities} 
In this section, we collect a few identities between the $C_{ij}$ and
$D_{ij}$ functions.  Their definition can be found in
Ref.~\cite{PV}.  Please note that we use a $(+,-,-,-)$ metric tensor, so that
Passarino-Veltman recurrence relations are the same as ours if we make the
substitution $p\cdot q\to -p\cdot q$ and $\de_{\mu\nu} \to -g_{\mu\nu}$ in
their formulae.
 
Starting with a three-point vertex function
\beq 
\int \frac{d^Dk}{i\pi^{D/2}}  
\frac{f(k)} 
{[k^2-\m^2][(k+p)^2-\m^2][(k+p+q)^2-\m^2]}, 
\eeq 
where $f(k)$ is an arbitrary function, that, for our purpose, will take the 
values $f(k)=1, \ k^\al,\ k^\al k^\bt$, and 
imposing the equality between this integral and the same integral where  
the integration variable has been shifted according to $k \to -k-p-q$, we 
have 
\beqn 
&& \mbox{}\hspace{-2cm}\int \frac{d^Dk}{i\pi^{D/2}}  
\frac{f(k)} 
{[k^2-\m^2][(k+p)^2-\m^2][(k+p+q)^2-\m^2]} = \nonumber\\ 
&&\int \frac{d^Dk}{i\pi^{D/2}}  
\frac{f(-k-p-q)} 
{[(k+p+q)^2-\m^2][(k+q)^2-\m^2][k^2-\m^2]}. 
\eeqn 
If $f(k)=1$, this identity gives  
\beq 
C_{0}(p,q) = C_{0}(q,p),  
\eeq 
while if $f(k)=k^\al$, it gives 
\beqn 
C_{11}(p,q) &=& -C_{12}(q,p)-C_{0}(q,p),\nonumber\\ 
C_{12}(p,q) &=& -C_{11}(q,p)-C_{0}(q,p)\;.
\eeqn 
In a similar way, if $f(k)=k^\al k^\bt$ we get 
\beqn 
C_{21}(p,q) &=& C_{22}(q,p)+2\,C_{12}(q,p)+C_{0}(q,p),\nonumber\\ 
C_{22}(p,q) &=& C_{21}(q,p)+2\,C_{11}(q,p)+C_{0}(q,p),\nonumber\\ 
C_{23}(p,q) &=& C_{23}(q,p)+C_{12}(q,p)+C_{11}(q,p)+C_{0}(q,p),\nonumber\\ 
C_{24}(p,q) &=& C_{24}(q,p). 
\eeqn 
Starting with a four-point function, we derive, in the same fashion,
\beqn 
D_{0}(p,q,l) &=& D_{0}(l,q,p),\nonumber\\ 
D_{11}(p,q,l) &=& -D_{13}(l,q,p)-D_{0}(l,q,p),\nonumber\\ 
D_{12}(p,q,l) &=& -D_{12}(l,q,p)-D_{0}(l,q,p),\nonumber\\ 
D_{13}(p,q,l) &=& -D_{11}(l,q,p)-D_{0}(l,q,p),\nonumber\\ 
D_{21}(p,q,l) &=& D_{23}(l,q,p)+2\,D_{13}(l,q,p)+D_{0}(l,q,p),\nonumber\\ 
D_{22}(p,q,l) &=& D_{22}(l,q,p)+2\,D_{12}(l,q,p)+D_{0}(l,q,p),\nonumber\\ 
D_{23}(p,q,l) &=& D_{21}(l,q,p)+2\,D_{11}(l,q,p)+D_{0}(l,q,p),\nonumber\\ 
D_{24}(p,q,l) &=& D_{26}(l,q,p)+D_{13}(l,q,p)+D_{12}(l,q,p)+D_{0}(l,q,p),
\nonumber\\ 
D_{25}(p,q,l) &=& D_{25}(l,q,p)+D_{13}(l,q,p)+D_{11}(l,q,p)+D_{0}(l,q,p),
\nonumber\\ 
D_{26}(p,q,l) &=& D_{24}(l,q,p)+D_{12}(l,q,p)+D_{11}(l,q,p)+D_{0}(l,q,p),
\nonumber\\ 
D_{27}(p,q,l) &=& D_{27}(l,q,p),\nonumber\\ 
D_{31}(p,q,l) &=&-3\,D_{13}(l,q,p)-3\,D_{23}(l,q,p)-D_{33}(l,q,p)-D_{0}(l,q,p), 
\nonumber\\ 
D_{32}(p,q,l) &=& -3\,D_{12}(l,q,p)-3\,D_{22}(l,q,p)-D_{32}(l,q,p)-D_{0}(l,q,p),
\nonumber\\ 
D_{33}(p,q,l) &=& -3\,D_{11}(l,q,p)-3\,D_{21}(l,q,p)-D_{31}(l,q,p)-D_{0}(l,q,p),
\nonumber\\ 
D_{34}(p,q,l) &=& -2\,D_{13}(l,q,p)-D_{12}(l,q,p)-2\,D_{26}(l,q,p)-D_{39}(l,q,p)
\nonumber\\ 
           && -D_{23}(l,q,p)-D_{0}(l,q,p),\nonumber\\ 
D_{35}(p,q,l) &=& -2\,D_{13}(l,q,p)-D_{11}(l,q,p)-2\,D_{25}(l,q,p)-D_{37}(l,q,p)
\nonumber\\ 
           && -D_{23}(l,q,p)-D_{0}(l,q,p),\nonumber\\ 
D_{36}(p,q,l) &=& -D_{13}(l,q,p)-2\,D_{12}(l,q,p)-2\,D_{26}(l,q,p)-D_{38}(l,q,p)
\nonumber\\ 
           && -D_{22}(l,q,p)-D_{0}(l,q,p),\nonumber\\ 
D_{37}(p,q,l) &=& -D_{13}(l,q,p)-2\,D_{11}(l,q,p)-2\,D_{25}(l,q,p)-D_{35}(l,q,p)
\nonumber\\ 
           && -D_{21}(l,q,p)-D_{0}(l,q,p),\nonumber\\ 
D_{38}(p,q,l) &=& -2\,D_{12}(l,q,p)-D_{11}(l,q,p)-2\,D_{24}(l,q,p)-D_{36}(l,q,p)
\nonumber\\ 
           && -D_{22}(l,q,p)-D_{0}(l,q,p),\nonumber\\ 
D_{39}(p,q,l) &=& -D_{12}(l,q,p)-2\,D_{11}(l,q,p)-2\,D_{24}(l,q,p)-D_{34}(l,q,p)
\nonumber\\ 
           && -D_{21}(l,q,p)-D_{0}(l,q,p),\nonumber\\ 
D_{310}(p,q,l) &=& -D_{13}(l,q,p)-D_{12}(l,q,p)-D_{11}(l,q,p)-D_{26}(l,q,p)
\nonumber\\ 
            && -D_{310}(l,q,p)-D_{25}(l,q,p)-D_{24}(l,q,p)-D_{0}(l,q,p),
\nonumber\\ 
D_{311}(p,q,l) &=& -D_{27}(l,q,p)-D_{313}(l,q,p),\nonumber\\ 
D_{312}(p,q,l) &=& -D_{27}(l,q,p)-D_{312}(l,q,p),\nonumber\\ 
D_{313}(p,q,l) &=& -D_{27}(l,q,p)-D_{311}(l,q,p). 
\eeqn

\section{Tensor and color structure of triangles} 
\label{app:triangle}
\begin{figure}[ht] 
   \centerline{\epsfig{file=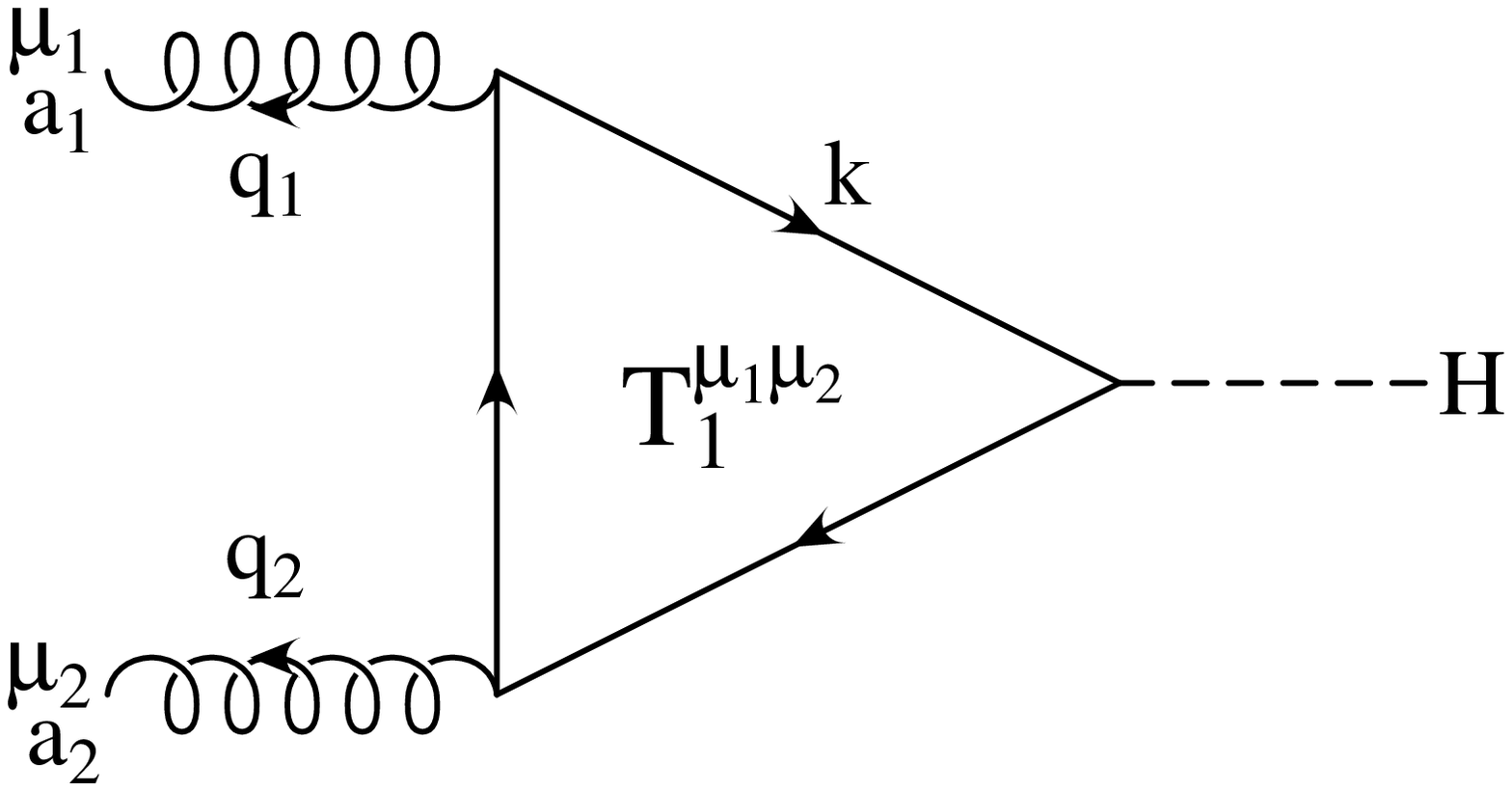,width=0.35\textwidth,clip=}  
               \ \ \ \    
               \epsfig{file=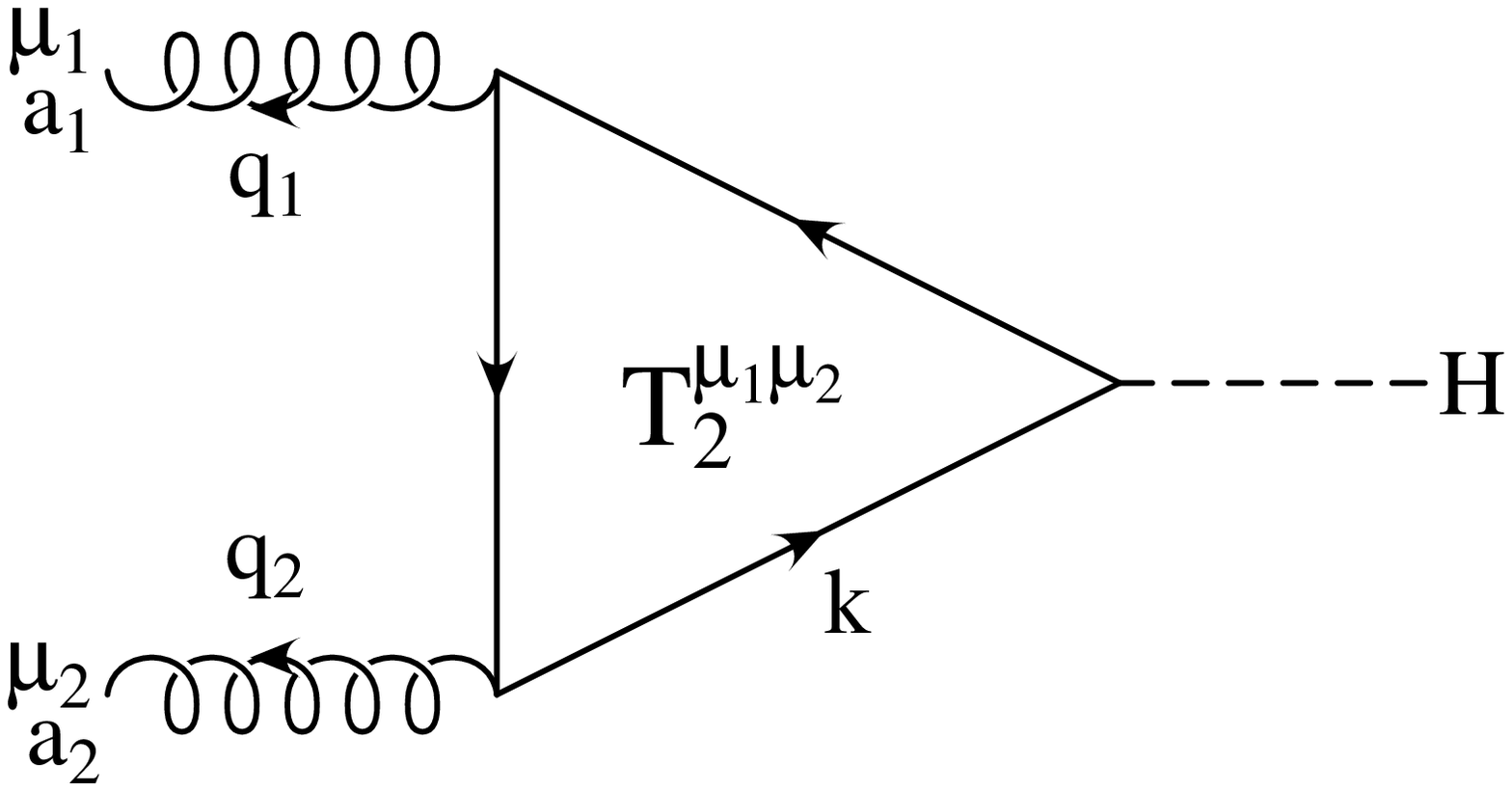,width=0.35\textwidth,clip=}} 
\ccaption{} 
{ \label{fig:tri_tens} Two three-point functions connected by charge 
               conjugation. }  
\end{figure} 
The two generic three-point functions
depicted in Fig.~\ref{fig:tri_tens} have the following expressions
\beqn 
T^{\mu_1\mu_2}_1(q_1,q_2) &=& {1\over 4\m}\int \frac{d^4k}{i\pi^{2}}  
\Tr{  \frac{1}{\sl{k}  -\m} \ga^{\mu_1} \frac{1}{\sl{k} + \sl{q}_1 -\m}
\ga^{\mu_2} \frac{1}{\sl{k} + \sl{q}_1 + \sl{q}_2 -\m}  },\\ 
T^{\mu_1\mu_2}_2(q_1,q_2) &=& {1\over 4\m}\int \frac{d^4k}{i\pi^{2}}  
\Tr{  \frac{1}{\sl{k}  -\m} \ga^{\mu_2} \frac{1}{\sl{k} + \sl{q}_2 -\m}
\ga^{\mu_1} \frac{1}{\sl{k} + \sl{q}_1+\sl{q}_2 -\m} }, 
\eeqn 
where $q_1$ and $q_2$ are outgoing momenta and where we put an
overall factor 
$1/(4\m)$ in front to delete a corresponding term coming from the trace over
the top quark. 
Using the charge-conjugation matrix $C$ 
\beq 
C \ga_\mu C^{-1} = - \ga^T_\mu, 
\eeq 
we can derive (Furry's theorem) 
\beq 
\label{eq:furry_tri}
  T^{\mu_1\mu_2}_1\(q_1,q_2\) = T^{\mu_1\mu_2}_2\(q_1,q_2\)
 \equiv T^{\mu_1\mu_2}\(q_1,q_2\). 
\eeq 
In addition,   
\beq 
\label{eq:p_dot_T} 
q_1^{\mu_1} T_{\mu_1\mu_2}(q_1,q_2) = q_2^{\mu_2} T_{\mu_1\mu_2}(q_1,q_2)  =0
\eeq 
expresses the gauge invariance of the triangle graphs. 
The generic tensor structure satisfying Eq.~(\ref{eq:p_dot_T}) is then
\beqn 
T^{\mu_1\mu_2}\(q_1,q_2\) &=& 
F_T\(q_1^2,q_2^2,\(q_1+q_2\)^2\) \, T_T^{\mu_1\mu_2}(q_1,q_2) +  
F_L\(q_1^2,q_2^2,\(q_1+q_2\)^2\) \, T_L^{\mu_1\mu_2}(q_1,q_2) \nonumber \\
&=& F_T \, T_T^{\mu_1\mu_2} + F_L \, T_L^{\mu_1\mu_2}, 
\eeqn 
where (dropping the dependence on external momenta $q_1$ and $q_2$,  
for ease of notation) 
\beqn 
T_T^{\mu_1\mu_2} &=& q_1\cdot q_2\, g^{\mu_1\mu_2} - q_1^{\mu_2}
\,q_2^{\mu_1}\;,  \\  
T_L ^{\mu_1\mu_2} &=&  q_1^2 \, q_2^2 \,g^{\mu_1\mu_2} 
- q_1^2 \,q_2^{\mu_1} \, q_2^{\mu_2} -  
q_2^2\, q_1^{\mu_1} \,q_1^{\mu_2} + q_1\cdot q_2 \, q_1^{\mu_1} q_2^{\mu_2}
\;, 
\eeqn 
and 
\beqn 
\label{eq:fl}   
F_L(q_1^2,q_2^2,Q^2) &=& -{1\over 2 \,\detw} 
\Bigg\{\lq 2-{3\,q_1^2\, q_2\cdot Q\over \detw}\rq 
	\Big(B_0\(q_1\)-B_0\(Q\)\Big) 
\nonumber\\&&  \phantom{-{1\over 2 \,\detw} }
+\lq 2-{3\,q_2^2\,q_1\cdot Q\over \detw}\rq 
	\Big(B_0\(q_2\)-B_0\(Q\)\Big)  
\nonumber\\&&\phantom{-{1\over 2 \,\detw} }
 -\lq 4\m^2+q_1^2+q_2^2+Q^2-{3\,q_1^2\,q_2^2\,Q^2\over \detw}\rq C_0\(q_1,q_2\)
+ R \Bigg\}\;,  \phantom{aaaaa} 
\\ 
\label{eq:ft} 
F_T(q_1^2,q_2^2,Q^2) &=& -{1\over 2\,\detw}
\bigg\{  
Q^2\Big[ B_0\(q_1\)+B_0\(q_2\)-2B_0\(Q\)-2\, q_1\cdot q_2\, C_0\(q_1,q_2\)\Big]
\nonumber \\ && 
\phantom{ -{1\over 2\,\detw} \bigg\{}
+\(q_1^2-q_2^2\) \Big( B_0\(q_1\)-B_0\(q_2\)\Big) \bigg\}  -q_1\cdot q_2 \,
F_L \;. 
\eeqn 
Here, the negative of $Q=q_1+q_2$ denotes the 
four-momentum of the Higgs boson,  
$\detw=q_1^2q_2^2-(q_1\cdot q_2)^2$ is the Gram determinant, and the terms   
proportional to $R=-2$ are pole residues in $D=4$ dimensions, originating 
from contributions proportional to $(D-4)\,C_{24}(q_1,q_2)$ in the tensor  
reduction procedure. 
Please note that even though the $B_0$ functions are divergent in 
$\ep=(4-D)/2$, the form factors $F_L$ and $F_T$ are finite.
 
If one of the external momenta, for example $q_1$, is light-like (real photon 
or gluon), with polarization vector $\ep_1$, then 
\beq 
T_L ^{\mu_1\mu_2} =   q_1^{\mu_1} \(  q_1\cdot q_2 \,  q_2^{\mu_2} -
q_2^2\,q_1^{\mu_2}  \)  
\quad  {\rm if} \quad  q_1^2 = 0, 
\eeq 
and as a consequence of the orthogonality $q_1 \cdot \ep_1=0$ we have 
\beq 
\label{eq:ep_T_L}	
\ep_{1\mu_1}\, T_L ^{\mu_1\mu_2} = 0\;, 
\eeq 
i.e.\ the form factor $F_L$ does not contribute when an on-shell gluon or
photon is  
attached to the triangle graph. 

The color structure of the sum of the two diagrams in Fig.~\ref{fig:tri_tens}
is straightforward. In fact, both the diagrams  have the same color
structure and we can write (see Eq.~(\ref{eq:furry_tri}))
\beq
\Tr{t^{a_1} t^{a_2}}  T^{\mu_1\mu_2}_1\(q_1,q_2\) +
\Tr{t^{a_2} t^{a_1}}  T^{\mu_1\mu_2}_2\(q_1,q_2\) =
\de^{a_1 a_2}  \, T^{\mu_1\mu_2}\(q_1,q_2\).
\eeq

\section{Tensor and color structure of boxes} 
\label{app:boxes}
\begin{figure}[ht]     
\centerline{\epsfig{file=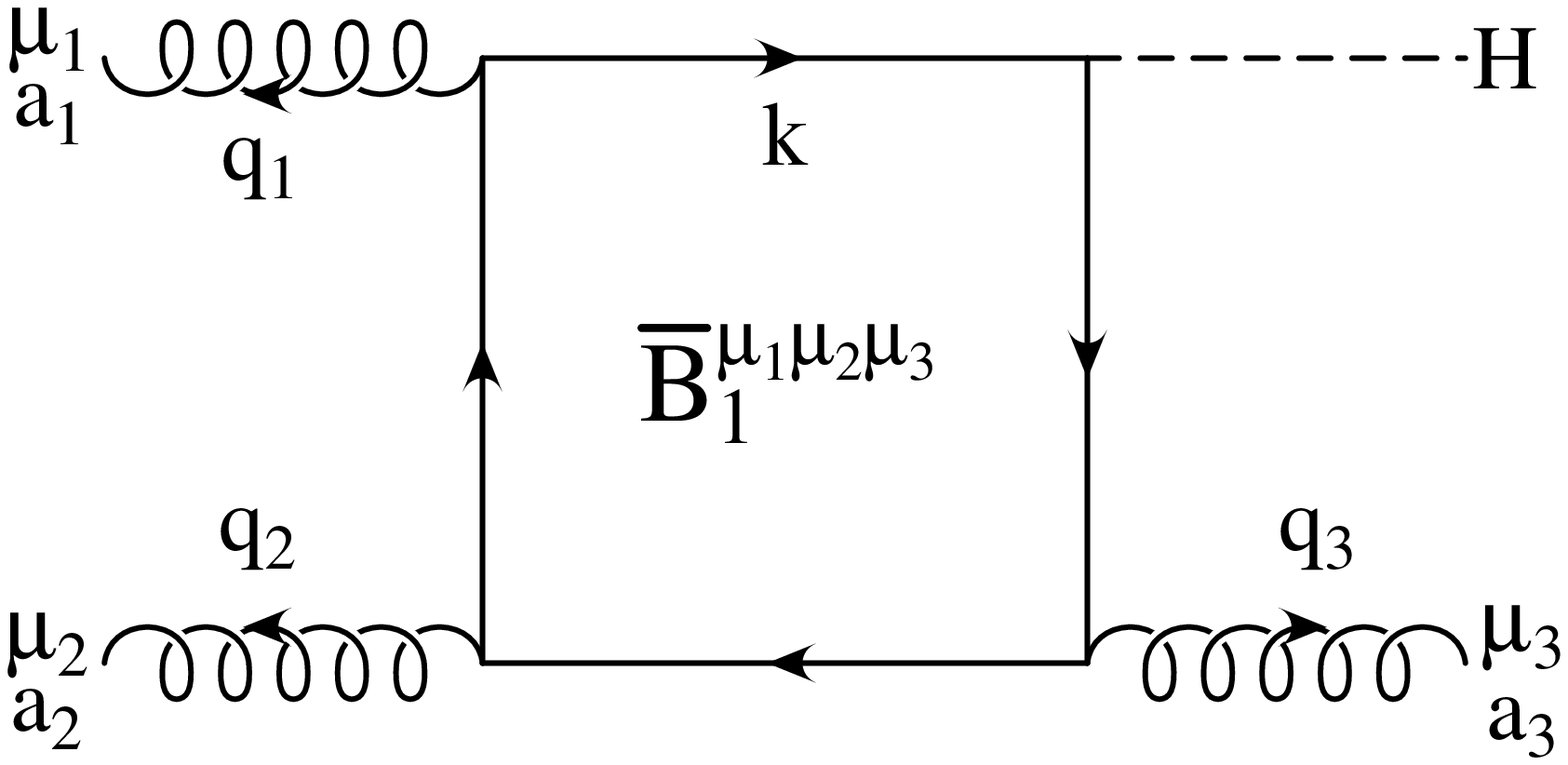,width=0.4\textwidth,clip=}  
               \ \ \ \  
               \epsfig{file=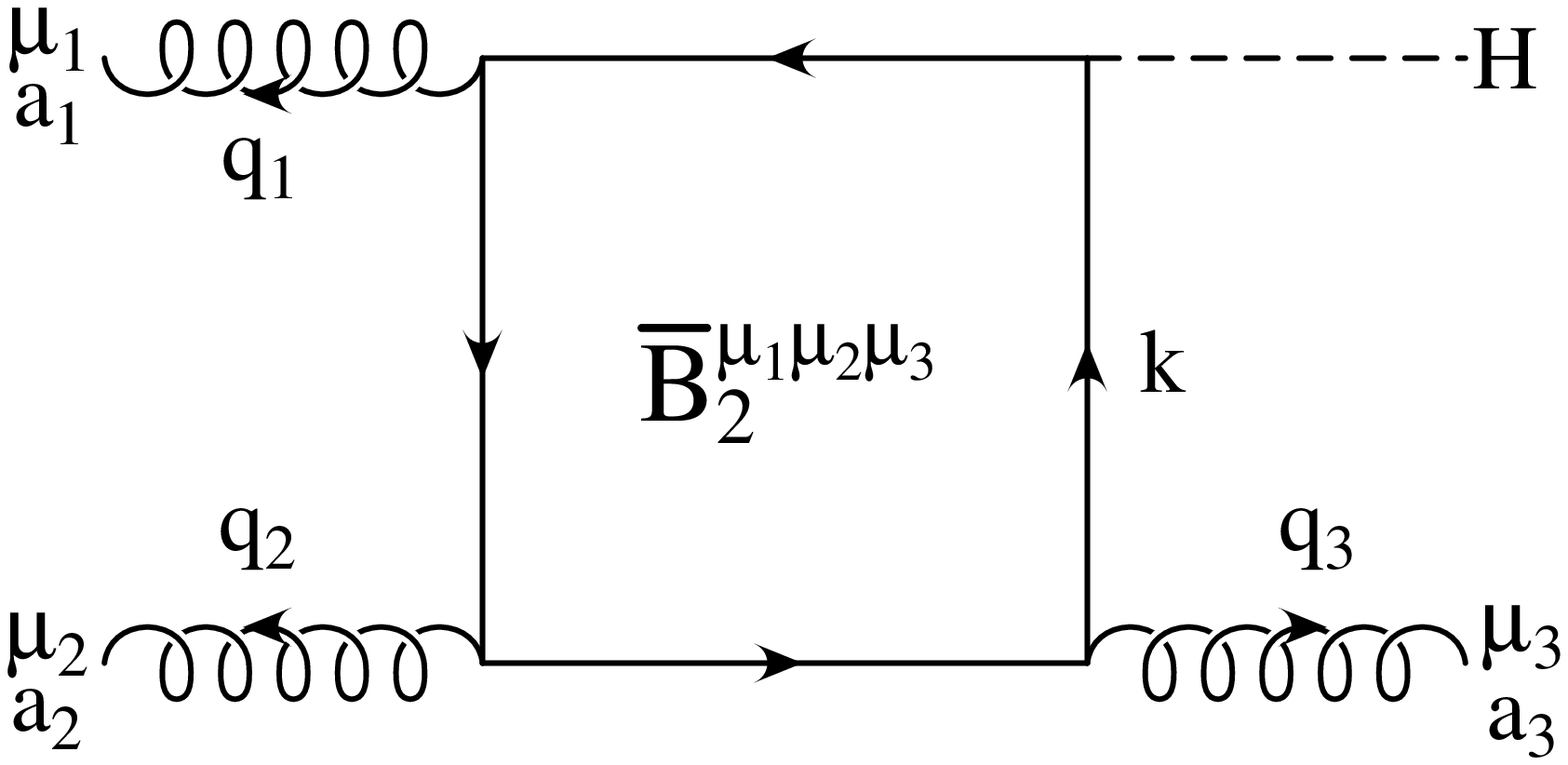,width=0.4\textwidth,clip=}} 
\ccaption{} 
{ \label{fig:ten_boxes} Two four-point functions connected by charge 
               conjugation.}  
\end{figure} 
The two generic four-point functions connected by charge conjugation, depicted
in Fig.~\ref{fig:ten_boxes}, have the following expressions
\beqn 
\overline{B}^{\mu_1\mu_2\mu_3}_1(q_1,q_2,q_3) \!\!&=&  \!\!
{1\over 4\m}\int \frac{d^4k}{i\pi^{2}} \,
{\rm Tr} \Bigg(
\frac{1}{\sl{k}  -\m} \ga^{\mu_1} \frac{1}{\sl{k} + \sl{q}_1 -\m}  \ga^{\mu_2} 
\nonumber \\ 
&& \phantom{{1\over 4\m}\int \frac{d^4k}{i\pi^{2}} \,
{\rm Tr} \Bigg(}
\times
\frac{1}{\sl{k} + \sl{q}_{12}  -\m}  \ga^{\mu_3}
\frac{1}{\sl{k} + \sl{q}_{123}  -\m} \Bigg),
\nonumber \\ 
\overline{B}^{\mu_1\mu_2\mu_3}_2(q_1,q_2,q_3)  \!\!&= \!\!& 
{1\over 4\m}\int \frac{d^4k}{i\pi^{2}} \,
{\rm Tr} \Bigg(
\frac{1}{\sl{k}  -\m}  \ga^{\mu_3} \frac{1}{\sl{k} + \sl{q}_3 -\m} \ga^{\mu_2} 
\nonumber \\ 
&& \phantom{ {1\over 4\m}\int \frac{d^4k}{i\pi^{2}} \,
{\rm Tr} \Bigg( }
\times 
\frac{1}{\sl{k} + \sl{q}_{23} -\m}  \ga^{\mu_1} 
\frac{1}{\sl{k} + \sl{q}_{123}  -\m} \Bigg), \phantom{aaaa}
\eeqn 
where $q_1, q_2$ and $q_3$ are the outgoing momenta, $q_{ij}=q_i+q_j$ and
$q_{ijk}=q_i+q_j+q_k$. The overall factor
$1/(4\m)$ cancels a corresponding term coming from the trace over
the top quark. 
From charge conjugation we have 
\beq 
\overline{B}^{\mu_1\mu_2\mu_3}_1(q_1,q_2,q_3) = 
- \overline{B}^{\mu_1\mu_2\mu_3}_2(q_1,q_2,q_3) \equiv  
\overline{B}^{\mu_1\mu_2\mu_3}(q_1,q_2,q_3)\;.
\eeq 
The color structure of the sum of the two diagrams depicted in
Fig.~\ref{fig:ten_boxes} is 
\beqn
\label{eq:final_col_box}
&&
\Tr{t^{a_1} t^{a_2} t^{a_3}} \overline{B}^{\mu_1\mu_2\mu_3}_1(q_1,q_2,q_3)
+
\Tr{t^{a_3} t^{a_2} t^{a_1}} \overline{B}^{\mu_1\mu_2\mu_3}_2(q_1,q_2,q_3)
\nonumber\\
&& \qquad\quad
= \lq \Tr{t^{a_1} t^{a_2} t^{a_3}} - \Tr{t^{a_3} t^{a_2} t^{a_1}} \rq
\overline{B}^{\mu_1\mu_2\mu_3}(q_1,q_2,q_3) = \frac{i}{2} f^{a_1 a_2 a_3}
\overline{B}^{\mu_1\mu_2\mu_3}(q_1,q_2,q_3),\nonumber\\
\eeqn
where we have used the identity
\beq
\Tr{t^{a_1} t^{a_2} t^{a_3}} = \frac{1}{4} \( d^{a_1 a_2 a_3} + i\, f^{a_1
a_2 a_3}\), 
\eeq
and the anti-symmetry of the structure constant $f^{a_1 a_2 a_3}$,
together with the symmetry of $d^{a_1 a_2 a_3}$.

The sum over the six gluon permutations of boxes is then proportional to the
single color factor $f^{a_1a_2a_3}$, and because of Bose symmetry of the
gluons, the kinematic box factor 
\bq
\label{eq:sum_six_boxes}
B^{\mu_1\mu_2\mu_3}(q_1,q_2,q_3) = 
\overline{B}^{\mu_1\mu_2\mu_3}(q_1,q_2,q_3) + 
\overline{B}^{\mu_2\mu_3\mu_1}(q_2,q_3,q_1) + 
\overline{B}^{\mu_3\mu_1\mu_2}(q_3,q_1,q_2) 
\eq
is totally antisymmetric in the gluon indices $(q_i,\mu_i)$, $i=1,2,3$.

\subsection{\boldmath $qg\,\to\,qgH$ and $gg\,\to\,ggH$}
The general structure of Eq.~(\ref{eq:sum_six_boxes}) can be further
restricted for the processes we are investigating:
$qg\,\to\,qgH$ and $gg\,\to\,ggH$. In fact, in both processes, two gluons of
the box are on-shell, and
the amplitude is contracted by the two corresponding polarization vectors
$\ep_i$, while the third gluon is contracted with the conserved current 
$J_{21}^{\mu_3}$ (see Eq.~(\ref{eq:J21})), in  the $qg\,\to\,qgH$ process,
and with the conserved current of the two on-shell gluons in the three-gluon
vertex, in  the $gg\,\to\,ggH$ case.

This gives rise to a few simplification in the structure of
Eq.~(\ref{eq:sum_six_boxes}).  In fact, a parity even, three-index tensor which
depends on three independent momenta (here taken as the three outgoing gluon
momenta $q_i,\;i=1,2,3$) can be written in terms of 36 independent tensor
structures, 9 of type $g^{\mu_1\mu_2}q_i^{\mu_3}$ and permutations, plus 27
tensors of type $q_i^{\mu_1}q_j^{\mu_2}q_k^{\mu_3}$. However, any terms
proportional to $q_1^{\mu_1}$, $q_2^{\mu_2}$, or $q_3^{\mu_3}$ vanish by
virtue of the transversity of the gluon polarization vectors $\eps_i^{\mu_i}$
and because the current on the vertex $\mu_3$ is conserved. 
This leaves us with 14 possible tensor structures, that can be further
reduced to three once we impose the total antisymmetry in the gluon indices
$(q_i,\mu_i)$
\beqn
\label{eq:boxtensor14}
B^{\mu_1\mu_2\mu_3} &=& 
g^{\mu_1\mu_2}q_1^{\mu_3}\; B_a(q_1,q_2,q_3) +
g^{\mu_2\mu_3}q_2^{\mu_1}\; B_a(q_2,q_3,q_1) +
g^{\mu_3\mu_1}q_3^{\mu_2}\; B_a(q_3,q_1,q_2) 
\nonumber \\{} &-& 
g^{\mu_2\mu_1}q_2^{\mu_3}\; B_a(q_2,q_1,q_3) -
g^{\mu_1\mu_3}q_1^{\mu_2}\; B_a(q_1,q_3,q_2) -
g^{\mu_3\mu_2}q_3^{\mu_1}\; B_a(q_3,q_2,q_1) 
\nonumber \\{} &+& 
q_3^{\mu_1}q_3^{\mu_2}q_1^{\mu_3}\;B_b(q_1,q_2,q_3) +
q_2^{\mu_1}q_1^{\mu_2}q_1^{\mu_3}\;B_b(q_2,q_3,q_1) +
q_2^{\mu_1}q_3^{\mu_2}q_2^{\mu_3}\;B_b(q_3,q_1,q_2)
\nonumber \\{} &-&
q_3^{\mu_1}q_3^{\mu_2}q_2^{\mu_3}\;B_b(q_2,q_1,q_3) -
q_2^{\mu_1}q_1^{\mu_2}q_2^{\mu_3}\;B_b(q_1,q_3,q_2) -
q_3^{\mu_1}q_1^{\mu_2}q_1^{\mu_3}\;B_b(q_3,q_2,q_1)
\nonumber \\{} &+&
q_2^{\mu_1}q_3^{\mu_2}q_1^{\mu_3}\;B_c(q_1,q_2,q_3) -
q_3^{\mu_1}q_1^{\mu_2}q_2^{\mu_3}\;B_c(q_2,q_1,q_3). 
\eeqn
Note that Bose symmetry implies that $B_c$ must be invariant under cyclic 
permutations of its arguments 
\beq
B_c(q_1,q_2,q_3) = B_c(q_2,q_3,q_1) = B_c(q_3,q_1,q_2).
\eeq
However, a convenient choice of gauge will remove the $B_c$ terms altogether,
as shown in Sec.~\ref{sec:qqgg}, 
using the polarization vectors defined in Eqs.~(\ref{eq:pol_vec1})
and~(\ref{eq:pol_vec2}). 

The scalar functions appearing in Eq.~(\ref{eq:boxtensor14}) are given by
\beqn
B_a(q_1,q_2,q_3) &=& 
{1\over 2}q_2\cdot q_3
\Bigl[ D_0(q_1,q_2,q_3) + D_0(q_2,q_3,q_1)+D_0(q_3,q_1,q_2) \Bigr] 
\nonumber \\
&& {}
-q_1\cdot q_2 \Bigl[   
D_{13}(q_2,q_3,q_1)+D_{12}(q_3,q_1,q_2)-D_{13}(q_3,q_2,q_1) \Bigr] 
- C_0(q_1,q_2+q_3) 
\nonumber\\
&&{} -
 4\Bigl[ D_{313}(q_2,q_3,q_1)+D_{312}(q_3,q_1,q_2)-D_{313}(q_3,q_2,q_1)
 \Bigr] \;,
\\
B_b(q_1,q_2,q_3) &=& 
D_{13}(q_1,q_2,q_3)+D_{12}(q_2,q_3,q_1)-D_{13}(q_2,q_1,q_3)
\nonumber\\
&&{}
 +4 \Bigl[ D_{37}(q_1,q_2,q_3)+D_{23}(q_1,q_2,q_3)+D_{38}(q_2,q_3,q_1)
\nonumber\\
&&{}
 + D_{26}(q_2,q_3,q_1)-D_{39}(q_2,q_1,q_3)-D_{23}(q_2,q_1,q_3)\Bigr]\;,
\\
B_c(q_1,q_2,q_3) &=& 
-{1\over 2}\Bigl[ D_0(q_1,q_2,q_3)+D_0(q_2,q_3,q_1)+D_0(q_3,q_1,q_2)\Bigr]
\nonumber\\
&&{}
 +4 \Bigl[ D_{26}(q_1,q_2,q_3)+D_{26}(q_2,q_3,q_1)+D_{26}(q_3,q_1,q_2)
\nonumber\\
&&{}
 + D_{310}(q_1,q_2,q_3)+D_{310}(q_2,q_3,q_1)+D_{310}(q_3,q_1,q_2)\Bigr]\;.
\eeqn

Further simplifications appear in the evaluation of $qg\,\to\, qgH$ matrix
elements.  In fact, since the polarization vectors for the two on-shell
gluons are either proportional to $x^\mu$ or $y^\mu$ (see
Eqs.~(\ref{eq:pol_vec1}) and~(\ref{eq:pol_vec2})), we only need the
contractions
$B_{xx\mu_3}=x_{\mu_1}x_{\mu_2}{B^{\mu_1\mu_2}}_{\mu_3}(q_1,q_2,q_3)$,
$B_{xy\mu_3}=x_{\mu_1}y_{\mu_2}{B^{\mu_1\mu_2}}_{\mu_3}$, 
etc.
The $\mu_3$ index will be contracted with the fermion current $J_{21}^{\mu_3}$
(see Eq.~(\ref{eq:amp_qg})). Since $q_1$, $q_2$, $q_3$ and $y$ span 
Minkowski space, $J_{21}$ can be expanded as 
\bq
\label{eq:J21_no_q3}
J_{21}^\mu ={1\over \detx}\Bigl( q_1\cdot J_{21}\;u^\mu
+q_2\cdot J_{21}\;v^\mu + y\cdot J_{21}\;y^\mu\Bigr)\;,
\eq
where 
\beqn
u^\mu &=& q_2\cdot q_3\;x^\mu + {\detx\over q_1\cdot q_2}\,q_2^\mu\;, \\
v^\mu &=& q_1\cdot q_3\;x^\mu + {\detx\over q_1\cdot q_2}\,q_1^\mu\;, 
\eeqn
and they satisfy the orthogonality relations
\beq
\begin{array}{llll}
u\cdot q_1 = \detx \;, \qquad &  u\cdot q_2=0\;, \qquad & u\cdot
q_3=0\;,\qquad  & u\cdot y = 0 \\
v\cdot q_1 = 0\;, \qquad & v\cdot q_2=\detx \;, \qquad & v\cdot
q_3=0\;,\qquad & v\cdot y = 0 \;,
\end{array}
\eeq
by virtue of the two on-shell conditions $q_1^2=0$ and $q_2^2=0$.
Note that there is no $q_3$ contraction in Eq.~(\ref{eq:J21_no_q3}), since
$q_3\cdot J_{21}=0$ by current conservation. 

The orthogonality of $y^\mu$ to all gluon momenta $q_i,\;i=1,2,3$, implies
that all contractions of $B^{\mu_1\mu_2\mu_3}$ with an 
odd number of $y$ will vanish. This leaves us with six non-zero contractions
of the tensor box integrals, $B_{yyu}$, $B_{yyv}$, $B_{yxy}$, $B_{xyy}$,
$B_{xxu}$ and $B_{xxv}$. Via the Bose symmetry of the tensor integral
in Eq.~(\ref{eq:boxtensor14}), the first four and the last two are
related by a permutation of gluon momenta:
\beqn 
\label{eq:B_yyu}
B_{yyu}  &=& \(\detx\)^2\; B_a(q_1,q_2,q_3) \;,
\\
B_{yyv}&=& -\(\detx\)^2\; B_a(q_2,q_1,q_3) 
\;,\\
B_{yxy}&=& -{\(\detx\)^2\over q_1\cdot q_2}\; B_a(q_3,q_1,q_2) 
\;,\\
B_{xyy}&=& {\(\detx\)^2\over q_1\cdot q_2}\; B_a(q_3,q_2,q_1) 
\;,\\
B_{xxu} &=&
\(\detx\)^2\Bigg\{ B_a\(q_1,q_2,q_3\)-{q_2\cdot q_3\over q_1\cdot q_2}
\Big[ B_a\(q_3,q_1,q_2\)-B_a\(q_3,q_2,q_1\) \Big] \nonumber\\
&& \phantom{\(\detx\)^2\Bigg\{}
+{\detx \over (q_1\cdot q_2)^2} B_b\(q_1,q_2,q_3\) \Bigg\}
\nonumber \\
&=&-{(\detx)^2\over q_1\cdot q_2} \Biggl\{ (q_1\cdot q_2)^2 \Bigl[  
D_{13}(q_2,q_3,q_1)+D_{12}(q_3,q_1,q_2)-D_{13}(q_3,q_2,q_1)
\Bigr]
\nonumber \\ 
&&{}-{1\over 2} q_1\cdot q_2 \, q_2\cdot q_3 \Bigl[ 
D_0(q_1,q_2,q_3)+D_0(q_3,q_1,q_2) \Bigr]
-q_2\cdot q_3 \, q_2\cdot \(q_3+{q_1\over 2}\) D_0(q_2,q_3,q_1) 
\nonumber \\
&& {} 
+ 4\,q_1\cdot q_2  
\Bigl[ D_{313}(q_2,q_3,q_1)+D_{312}(q_3,q_1,q_2)-D_{313}(q_3,q_2,q_1) \Bigr] 
\nonumber\\
&&{}
+\Bigl[q_2\cdot q_3 \, q_3\cdot (q_1-q_2)-q_3^2 \, q_1\cdot q_2\Bigr]
 \Bigl[  D_{13}(q_1,q_2,q_3)+D_{12}(q_2,q_3,q_1)
        -D_{13}(q_2,q_1,q_3) \Bigr] 
\nonumber\\
&&{}
-4 \, q_2\cdot q_3 \Bigl[ 2\Bigl( 
D_{313}(q_1,q_2,q_3)+D_{312}(q_2,q_3,q_1)-D_{313}(q_2,q_1,q_3) 
\Bigr) +D_{27}(q_2,q_3,q_1)    \Bigr]
\nonumber\\
&&{}  -4 \,\detxt \Bigl[D_{37}(q_1,q_2,q_3)+D_{23}(q_1,q_2,q_3) +
D_{38}(q_2,q_3,q_1)   + D_{26}(q_2,q_3,q_1)
\nonumber\\
&&{}
-D_{39}(q_2,q_1,q_3)-D_{23}(q_2,q_1,q_3)\Bigr]
+ q_1\cdot q_2\, C_0(q_1,q_2+q_3) 
 \Biggr\} \;,
\\
\label{eq:B_xxv}
B_{xxv} &=& -(\detx)^2\Bigg\{ B_a(q_2,q_1,q_3)+{q_1\cdot q_3\over 
q_1\cdot q_2} \Big[ B_a(q_3,q_1,q_2)-B_a(q_3,q_2,q_1) \Big]
\nonumber\\
&& \phantom{-(\detx)^2\Bigg\{}
+{\detx \over (q_1\cdot q_2)^2}B_b(q_2,q_1,q_3) \Bigg\}
\nonumber \\
&=& -B_{xxu}(q_1 \leftrightarrow q_2) \;.
\eeqn 
Note that in the replacement $B_{xxv}=-B_{xxu}(q_1 \leftrightarrow q_2)$,
 the Gram determinant, $\detx$, is to 
be treated as totally symmetric under interchange of gluon momenta 
$q_1$, $q_2$, and $q_3$.

\section{Tensor and color structure of pentagons} 
\label{app:pentagons}
\begin{figure}[ht]     
\centerline{\epsfig{file=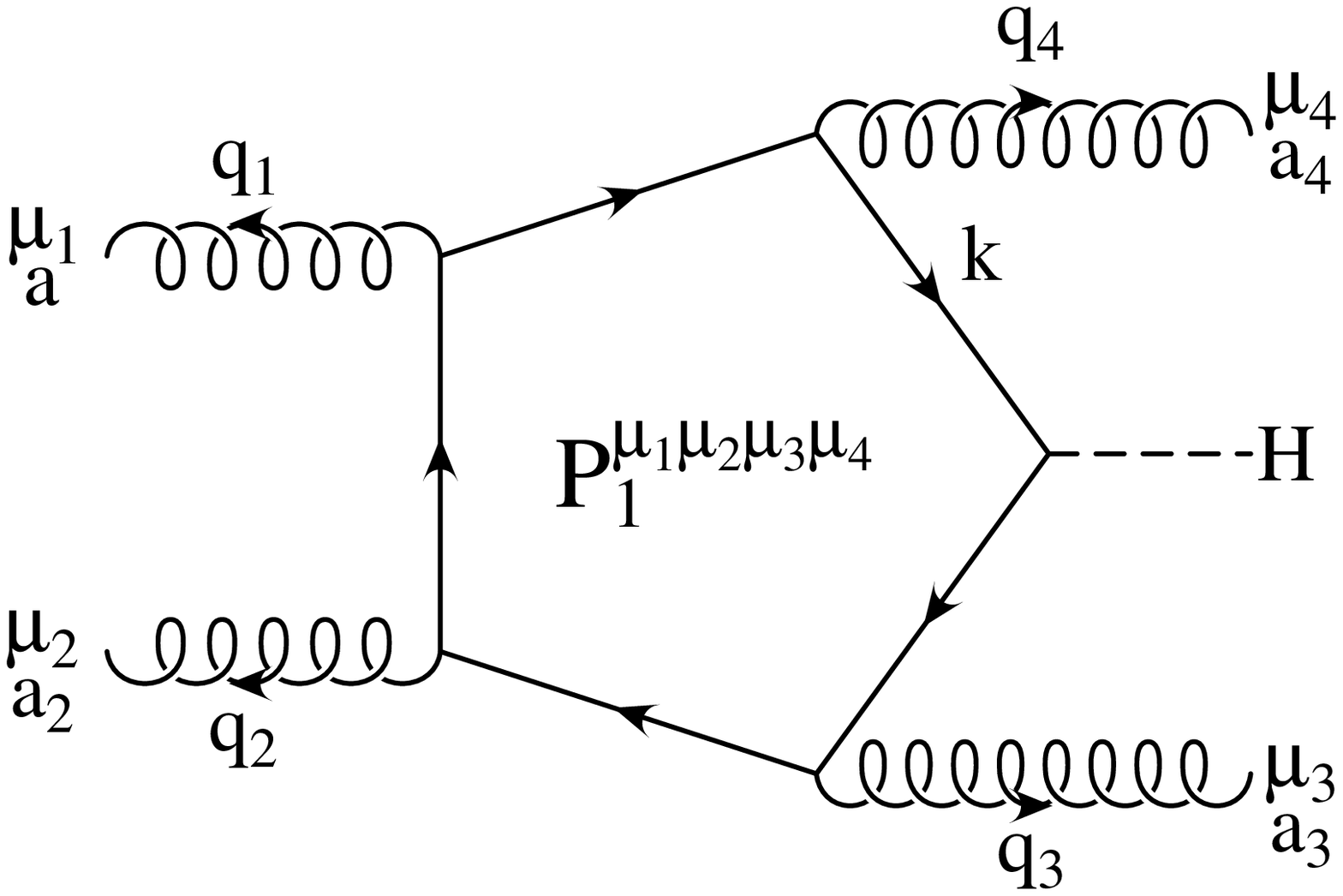,width=0.4\textwidth,clip=}  
               \ \ \ \  
               \epsfig{file=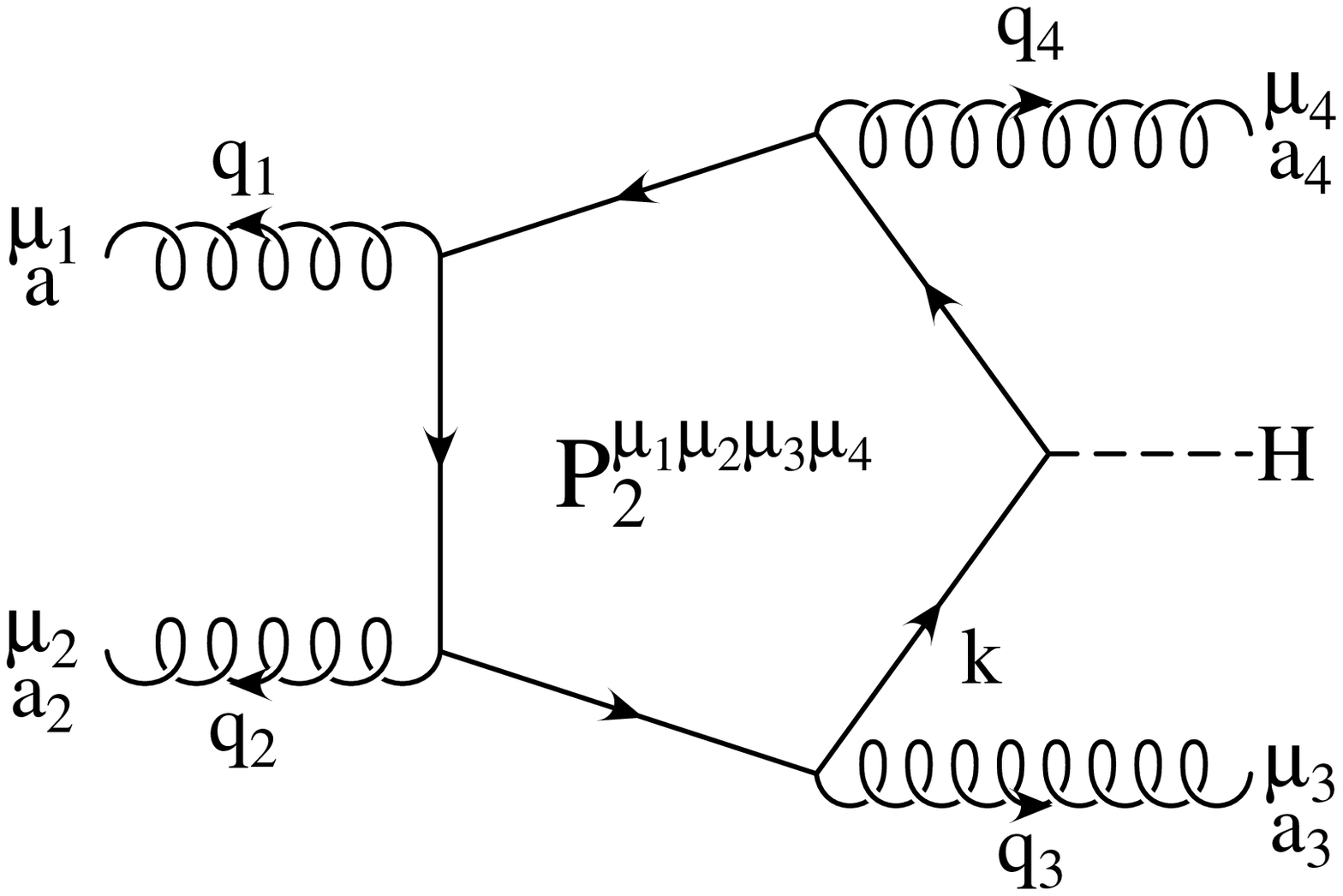,width=0.4\textwidth,clip=}} 
\ccaption{} 
{ \label{fig:ten_pentagons} Two five-point functions connected by charge 
               conjugation.}  
\end{figure} 
The two generic five-point functions connected by charge conjugation depicted
in Fig.~\ref{fig:ten_pentagons} have the following expression
\beqn 
P^{\mu_1\mu_2\mu_3\mu_4}_1(q_1,q_2,q_3,q_4) \!\!&=&  \!\!
{1\over 4\m}\int \frac{d^4k}{i\pi^{2}} \,
{\rm Tr} \Bigg(
\frac{1}{\sl{k}   -\m}  \ga^{\mu_4}
\frac{1}{\sl{k} + \sl{q}_4 -\m} \ga^{\mu_1} 
\frac{1}{\sl{k} + \sl{q}_{14} -\m}  \ga^{\mu_2} 
\nonumber \\ 
&& \phantom{{1\over 4\m}\int \frac{d^4k}{i\pi^{2}} \,
{\rm Tr} \Bigg(}
\times
\frac{1}{\sl{k} + \sl{q}_{124} -\m}  \ga^{\mu_3}
\frac{1}{\sl{k} + \sl{q}_{1234}  -\m}  
\Bigg),
\nonumber \\ 
P^{\mu_1\mu_2\mu_3\mu_4}_2(q_1,q_2,q_3,q_4) \!\!&=&  \!\!
{1\over 4\m}\int \frac{d^4k}{i\pi^{2}} \,
{\rm Tr} \Bigg(
\frac{1}{\sl{k}   -\m}  \ga^{\mu_3}
\frac{1}{\sl{k} + \sl{q}_3 -\m} \ga^{\mu_2} 
\frac{1}{\sl{k} + \sl{q}_{23} -\m}  \ga^{\mu_1} 
\nonumber \\ 
&& \phantom{{1\over 4\m}\int \frac{d^4k}{i\pi^{2}} \,
{\rm Tr} \Bigg(}
\times
\frac{1}{\sl{k} + \sl{q}_{123} -\m}  \ga^{\mu_4}
\frac{1}{\sl{k} + \sl{q}_{1234}  -\m}  
\Bigg), \phantom{aaaa}
\eeqn 
where $q_1, q_2, q_3$ and $q_4$ are the outgoing momenta ($q_{ij}=q_i+q_j$ and
similar ones for $q_{ijl}$ and $q_{ijln}$) 
and where we put an overall factor
$1/(4\m)$ in front to delete a corresponding term coming from the trace over
the top quark. 
From charge conjugation we have 
\beq 
{P}^{\mu_1\mu_2\mu_3\mu_4}_1(q_1,q_2,q_3,q_4) = 
{P}^{\mu_1\mu_2\mu_3\mu_4}_2(q_1,q_2,q_3,q_4) \equiv  
{P}^{\mu_1\mu_2\mu_3\mu_4}(q_1,q_2,q_3,q_4)\;.
\eeq 
Finally, the color structure of the sum of the two diagrams depicted in
Fig.~\ref{fig:ten_pentagons} is
\beqn
\label{eq:final_col_pent}
&&
\Tr{t^{a_1} t^{a_2} t^{a_3} t^{a_4}}
{P}^{\mu_1\mu_2\mu_3\mu_4}_1(q_1,q_2,q_3,q_4)  + 
\Tr{t^{a_4} t^{a_3} t^{a_2} t^{a_1}}
{P}^{\mu_1\mu_2\mu_3\mu_4}_2(q_1,q_2,q_3,q_4)  
\nonumber\\
&& \qquad\qquad
= \lq\Tr{t^{a_1} t^{a_2} t^{a_3} t^{a_4}} + 
\Tr{t^{a_1} t^{a_4} t^{a_3} t^{a_2}}
 \rq  {P}^{\mu_1\mu_2\mu_3\mu_4}(q_1,q_2,q_3,q_4)\;.
\eeqn
Further details about the color structure of pentagons are given in
Sec.~\ref{sec:gg_ggH}.


\end{document}